\documentclass[sigconf]{acmart}

\usepackage{graphicx}
\usepackage{latexsym}
\usepackage{amsmath}
\usepackage{amsthm}
\usepackage{booktabs}
\usepackage{algorithm}
\usepackage{algorithmic}
\usepackage{amsfonts}
\usepackage{xspace}
\usepackage{xcolor}
\usepackage{bbm}
\usepackage{subfig}
\usepackage[switch]{lineno}

\theoremstyle{definition}
\newtheorem{definition}{Definition}[section]

\newcommand{\app}{WeChat\xspace}
\newcommand{\model}{CMT\xspace}
\newcommand{\group}{WeChat chat group\xspace}

\newcommand{\ADD}{\texttt{ADD}\xspace}
\newcommand{\CREATE}{\texttt{CREATE}\xspace}
\newcommand{\ENTER}{\texttt{ENTER}\xspace}
\newcommand{\PULL}{\texttt{PULL}\xspace}
\newcommand{\POST}{\texttt{POST}\xspace}
\newcommand{\FINISH}{\texttt{FINISH}\xspace}
\newcommand{\SEND}{\texttt{SEND}\xspace}
\newcommand{\LOGIN}{\texttt{LOGIN}\xspace}
\newcommand{\TRANSFER}{\texttt{TRANSFER}\xspace}
\newcommand{\DISAPPEAR}{\texttt{DISAPPEAR}\xspace}

\begin{document}

\title{Crowdsourcing Fraud Detection over Heterogeneous Temporal MMMA Graph}

\author{Zequan Xu}
\affiliation{
 \country{School of Informatics, Xiamen University}
}
\email{xuzequan@stu.xmu.edu.cn}

\author{Qihang Sun}
\affiliation{
  \country{Tencent Inc.}
}
\email{aaronqhsun@tencent.com}

\author{Shaofeng Hu}
\affiliation{
  \country{Tencent Inc.}
}
\email{hugohu@tencent.com}

\author{Jieming Shi}
\affiliation{
  \country{The Hong Kong Polytechnic University}
}
\email{jieming.shi@polyu.edu.hk}

\author{Hui Li}
\authornote{Corresponding Author.}
\affiliation{
  \country{School of Informatics, Xiamen University}
}
\email{hui@xmu.edu.cn}

\begin{abstract}
The rise of the click farm business using Multi-purpose Messaging Mobile Apps (MMMAs) tempts cybercriminals to perpetrate crowdsourcing frauds that cause financial losses to click farm workers. In this paper, we propose a novel contrastive multi-view learning method named CMT for crowdsourcing fraud detection over the heterogeneous temporal graph (HTG) of MMMA. CMT captures both heterogeneity and dynamics of HTG and generates high-quality representations for crowdsourcing fraud detection in a self-supervised manner. We deploy CMT to detect crowdsourcing frauds on an industry-size HTG of a representative MMMA WeChat and it significantly outperforms other methods. CMT also shows promising results for fraud detection on a large-scale public financial HTG, indicating that it can be applied in other graph anomaly detection tasks. We provide our implementation at \url{https://github.com/KDEGroup/CMT}.
\end{abstract}

\maketitle

\section{Introduction}
\label{sec:intro}

\begin{figure}[t]
	\begin{center}
		\includegraphics[width=1\linewidth]{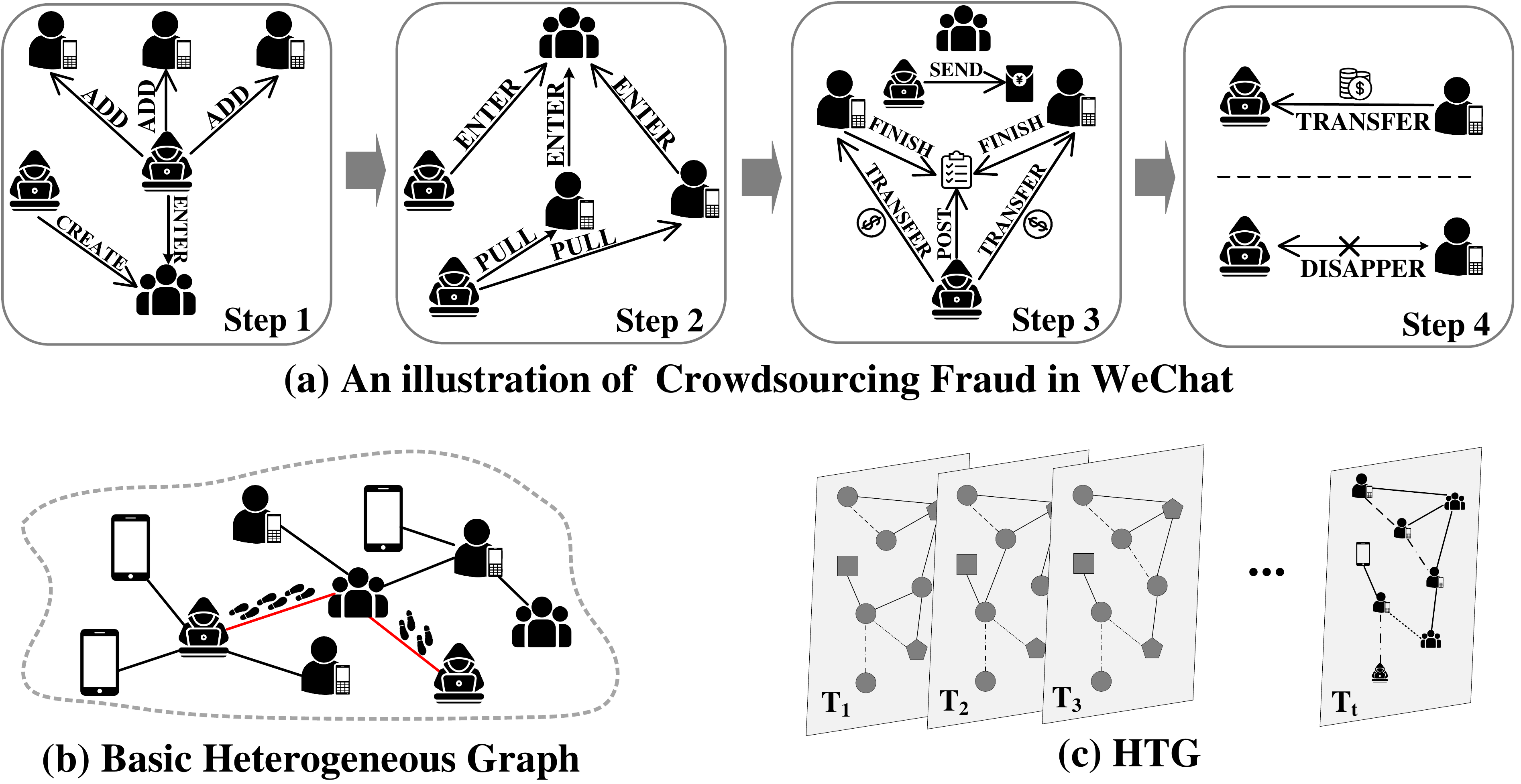}
	\end{center}
	\caption{Crowdsourcing fraud in \app.}
	\label{fig:illustration}
\end{figure}

\underline{M}ulti-purpose \underline{M}essaging \underline{M}obile \underline{A}pps (MMMAs) like \app\footnote{\url{https://www.wechat.com/en}} integrate several functionalities (e.g., chat, make transactions or book tickets) into one app, bringing great convenience and attracting billions of users~\cite{wechatuser}.
Due to the huge amount of users, MMMAs have become popular platforms for click farms~\cite{clickfarm} where \emph{click farm workers} are recruited to complete online tasks (e.g., click on YouTube videos to create appearance of online popularity).
MMMAs are used to hire workers, assign tasks and deliver rewards.
However, the rise of the click farm business tempts cybercriminals to perpetrate \emph{crowdsourcing frauds} where victims are click farm workers.

This paper studies the detection of crowdsourcing frauds and we get the permission to experiment on \app. 
Fig.~\ref{fig:illustration}(a) depicts a typical process of crowdsourcing fraud in \app: 
\begin{itemize}
	
	\item \textbf{Step 1:} Fraudsters send ``add friend'' requests to potential victims (i.e., \ADD in Fig.~\ref{fig:illustration}(a) - Step 1). Chat groups are created and cybercriminals join (i.e., \CREATE and \ENTER in Fig.~\ref{fig:illustration}(a) - Step 1). They disguise as normal users.
	
	\item \textbf{Step 2:} Fraudsters invite victims (i.e., \PULL in Fig.~\ref{fig:illustration}(a) - Step 2) to join fraud groups (i.e., \ENTER in Fig.~\ref{fig:illustration}(a) - Step 2) by using high reward as bait.

	\item \textbf{Step 3:} Group members are encouraged to complete tasks (i.e., \FINISH in Fig.~\ref{fig:illustration}(a) - Step 3) posted by the group owner (i.e., \POST in Fig.~\ref{fig:illustration}(a) - Step 3).  Typical tasks include transferring money to buy some products or top up online shopping cards. To gain victims' trust, the cost of the first a few tasks is not high and fraudsters pay the commission as promised (i.e., \TRANSFER in Fig.~\ref{fig:illustration}(a) - Step 3). 
	Before victims complete tasks, fraudsters may even send bonus packages in the group by making group transactions (i.e., \SEND in Fig.~\ref{fig:illustration}(a) - Step 3) as an incentive. And all group members can receive a random portion of the bonus.
	
	\item \textbf{Step 4:} With victims' guard down, fraudsters post new tasks that request much more money. 
	Victims can see that other group members (fraud conspirators) complete and get high rewards. 
	Hence, they are deceived and transfer money to complete new tasks (i.e., \TRANSFER in Fig.~\ref{fig:illustration}(a) - Step 4).
	However, after that, fraudsters disappear and do not response anymore  (i.e., \DISAPPEAR in Fig.~\ref{fig:illustration}(a) - Step 4).  
	
\end{itemize}

\vspace{5pt}
The social nature of MMMAs gives a natural motivation to model MMMA as a user-user interaction graph. 
We can observe that MMMA graph is both \emph{heterogeneous} and \emph{dynamic}: 
(1) As shown in Fig.~\ref{fig:illustration}(a), users can perform diverse operations due to MMMAs' all-in-one functionality.  
(2) Crowdsourcing frauds typically involve actions spanning multiple time points (e.g., Fig.~\ref{fig:illustration}(a) shows a crowdsourcing fraud spanning four time points). Hence, the MMMA graph can also be modeled as a dynamic graph~\cite{YouDL22}.
Moreover, crowdsourcing fraud detection \emph{lacks of supervision} since (1) it is hard to label the huge volume of MMMA users, and (2) user features are limited and accessing private information like chat content is forbidden in order to protect user privacy.

The heterogeneity and the dynamics of the MMMA graph require the detection approach to model user behaviors from \emph{multiple views} so that both diverse user interactions and temporal features can be fully leveraged to uncover trails of frauds. 
Moreover, the detection model should seek additional supervision to overcome the lack of supervision.
To solve the above issues, we propose \underline{C}ontrastive \underline{M}ul\underline{t}i-view Learning over Heterogeneous Temporal Graph (\model) for crowdsourcing fraud detection. 
Our contributions are: 
\begin{itemize}
	
	\item We design a Heterogeneous Temporal Graph (HTG) to model the MMMA data for crowdsourcing fraud detection. 
	
	\item We propose a novel method \model for crowdsourcing fraud detection. \model uses a heterogeneous graph encoder to capture the heterogeneity of HTG. To model dynamics of HTG, \model constructs two types of user history sequences as two ``views'' of behavior patterns. \model further augments each sequence and its contrastive learning encoder encodes sequences in a self-supervised manner. 
    
	\item We deploy CMT to detect crowdsourcing frauds on an industry-size HTG of WeChat and it significantly outperforms other methods. Additional promising experimental results on a large-scale financial HTG show that \model can also be applied in other graph anomaly detection (GAD) tasks like financial fraud detection.  
\end{itemize}

\section{Related Work}
\label{sec:related}

\subsection{Graph-based Anomaly Detection (GAD)} 
GAD detects anomalous graph objects (i.e., nodes, edges or sub-graphs)~\cite{9565320}. Early methods mainly use handcrafted feature engineering or statistical models built by human experts~\cite{LiSCGY14}, which is difficult to generalize. 
Recent works are mostly inspired by deep learning techniques.  
DOMINANT~\cite{DingLBL19} utilizes GCN for graph representation learning and leverages the learned embeddings to reconstruct the original adjacent matrix for anomaly detection. 
ALARM~\cite{9162509} further employs multiple attributed views to describe different perspectives of the objects for anomalies detection. 
CARE-GNN~\cite{DouL0DPY20} uses label-aware similarity measure and reinforcement learning to find informative neighboring nodes for aggregation to avoid the impact of camouflage behaviors of fraudsters.
PC-GNN~\cite{LiuAQCFYH21} adopts various samplers for imbalanced supervised learning GAD. 
Different from previous methods that jointly learn node representation and the classifier, DCI~\cite{Wang0GYL021}, inspired by the recent advances of self-supervised learning, decouples these two phases for GAD. 

\subsection{Anomaly Detection in Dynamic Graphs}
Anomaly detection in dynamic graphs attracts increasing interest,  
since many real-world networks can be generally represented in the form of dynamic graphs.  
Early methods such as CAD~\cite{SricharanD14} detect node relationships responsible for abnormal changes in graph structure by tacking a measure that combines information regarding changes in both graph structure and edge weights. StreamSpot~\cite{ManzoorMA16}  
uses an online centroid-based clustering and anomaly detection scheme to evaluate the anomalousness of the incoming graphs. 
Another branch of approaches employs deep learning technique. Zheng et al.~\cite{ZhengLLLG19} first utilizes temporal GCN and attention mechanism to model short-term and long-term patterns. Then a GRU network is introduced to process such patterns and encode temporal features. 
NetWalk~\cite{YuCAZCW18} adopts a random walk based encoder to learn the network representations 
and employs a clustering-based anomaly detector to score the abnormality of each edge.  
StrGNN~\cite{CaiCLGN0C21}  
extracts the $h$-hop enclosing subgraph of edges and labels each node to identify its corresponding role in the subgraph. Then, it leverages GCN and GRU to capture the spatial and temporal information for anomaly detection.

\section{Our Framework \model}
\label{sec:method}

\subsection{Overview}

\begin{figure*}[!t]
	\begin{center}
		\includegraphics[width=1\linewidth]{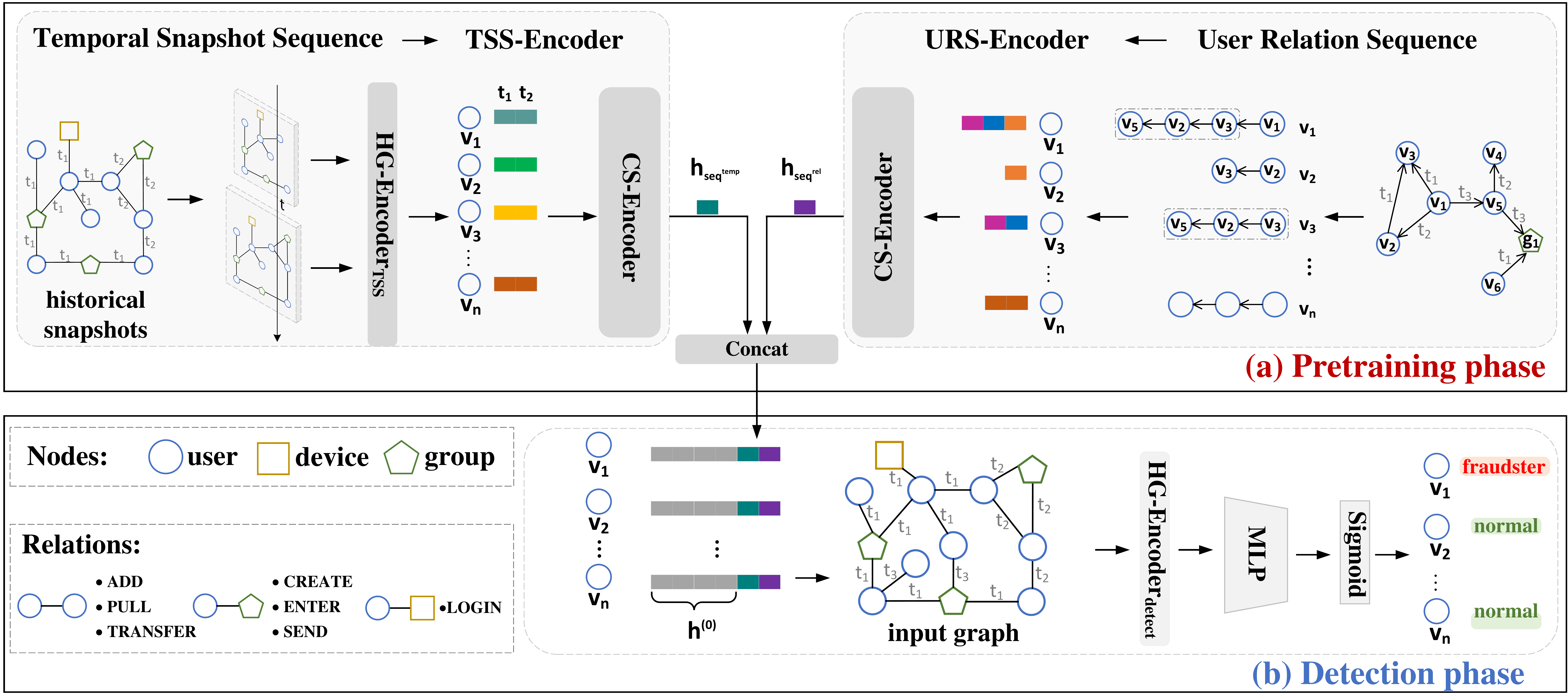}
	\end{center}
	\caption{Overview of \model.}
	\label{fig:overview}
\end{figure*}

Fig.~\ref{fig:overview} provides an overview of \model.
We model the MMMA graph as a \underline{H}eterogeneous \underline{T}emporal \underline{G}raph (HTG, Sec.~\ref{sec:graph_construct}).
\model detects crowdsourcing frauds over the HTG and it consists of a pretraining phase and a detection phase. 
In the pretraining phase, \model learns to model both graph heterogeneity and dynamics: 
(1) \model firstly uses a Heterogeneous GNN Encoder (HG-Encoder) to encode heterogeneity (Sec.~\ref{sec:graph});
(2) For graph dynamics, we construct two types of user history sequence as two ``views'' of user behavior patterns (Sec.~\ref{sec:user_seq});
(3) Then, we design data augmentation methods and generate two augmented sequences for each user history sequence (Sec.~\ref{sec:data_aug}) to provide additional supervision to enhance \model's understanding of the graph; 
(4) After that, \model adopts two contrastive learning enhanced Transformer encoders (CS-Encoder), namely TSS-Encoder and URS-Encoder, to encode user history sequences from different views (Sec.~\ref{sec:seq_encoder}). 
In the detection phase, \model estimates the abnormality of each user. 

\subsection{Modeling MMMA Graph as HTG}
\label{sec:graph_construct}

\begin{table}[t]
	\centering
	\caption{Description for binary node attributes in \app dataset}
	\label{tab:feature}
	\scalebox{0.9}{
		\begin{tabular}{|l|l|}
			\hline
			\textbf{ID} & \multicolumn{1}{c|}{\textbf{Description}}                                                                      \\ \hline
			1           & Was the account registered in the past 90 days?                                                                \\ \hline
			2           & Is the account currently active?                                                                               \\ \hline
			3           & Has the account owner verified his/her real identity?                                                          \\ \hline
			4           & Has the account owner changed his/her real identity?                                                           \\ \hline
			5           & \begin{tabular}[c]{@{}l@{}}Is the account currently banned by any app or \\ third-party services?\end{tabular} \\ \hline
			6           & Was the account reported by other users in the past 14 days?                                                   \\ \hline
			7           & Is the account rented?                                                                                         \\ \hline
		\end{tabular}
	}
\end{table}

Firstly, we pre-extract some features $\mathbf{p}_u\in \mathbb{R}^{7}$ for each MMMA user $u$ in \app. 
\emph{Note that they are chosen through a strict investigation process to protect users' privacy}.
Tab.~\ref{tab:feature} describes 7 node attributes used as features.
For each node, we construct a binary vector where 0/1 in each dimension indicates the existence/non-existence of an attribute:
\begin{itemize}
	\item Attributes 1 - 4 describe the ``value'' of \app accounts: accounts that are registered early, currently active or having verified and consistent identity are more valuable.
	\item Attributes 5 - 7 are related to detect crowdsourcing fraud, but they are common features across different anomaly detection tasks in \app. For example, (1) normal businessmen also rent multiple accounts to advertise their products in chat groups, and (2) an account can be reported by other users or banned in some services (e.g., games) due to various reasons like unfriendly behavior in games, viral marketing or verbal abuse. Hence, only considering these attributes is not sufficient for detecting crowdsourcing fraudsters accurately.
\end{itemize}

Then, we construct a HTG to model MMMA data. 
We first consider a basic heterogeneous graph (depicted in Fig.~\ref{fig:illustration}(b)) with three node types (i.e., user, group and device). 
A device node indicates the device that a user uses to login, and it is uniquely identified by IP address and device type.
We include seven relation types: ``create a group'' (\CREATE), ``join a group'' (\ENTER),  ``login on a device'' (\LOGIN), ``invite someone to join a group'' (\PULL), ``send a bonus packet to a group'' (\SEND), ``become \app friends'' (\ADD), ``transfer money'' (\TRANSFER). 
Based on the basic graph, we further consider temporal dependencies (i.g., evolving user states and behaviors) and construct a HTG as shown in Fig.~\ref{fig:illustration}(c):
\begin{definition}[HTG] A HTG is a graph stream consists of discrete snapshots. Let the latest timestamp be $T$. A graph stream can be denoted as $\{G^t\}_{t=1}^T$, where each $G^t=(V^t,E^t)$ is a basic heterogeneous graph at time point $t$. 
\end{definition}

\subsection{Graph Heterogeneity Encoding} 
\label{sec:graph}

For simplicity, in the following, we use the basic heterogeneous graph to illustrate the Heterogeneous GNN Encoder (HG-Encoder) for encoding graph heterogeneity in \model.

We first project a user $u$'s raw features $\mathbf{p}_u$ and utilize the projection as the initial embedding for $u$: $\mathbf{h}_u^{(0)} = \mathbf{W}_{\mathbf{h}}\mathbf{p}_u$, where $\mathbf{W}_{\mathbf{h}}$ is a learnable matrix.  
For each group/device node, we aggregate embeddings of its members (i.e., user neighbors) as its initial embeddings:
\begin{equation}
	\label{eq:project}
	\mathbf{h}_g^{(0)} = \text{mean}(\{\mathbf{h}_{v},\forall v\in N_{g}\}),\,\,\mathbf{h}_d^{(0)} = \text{mean}(\{\mathbf{h}_{v},\forall v\in N_{d}\}),
\end{equation}
where $\mathbf{h}_g^{(0)}$ and $\mathbf{h}_d^{(0)}$ are initial embeddings for the group node $g$ and the device node $d$, respectively. $N_g$ and $N_d$ denotes neighboring user nodes of $g$ and $d$ in the HTG, respectively. $mean(\cdot)$ indicates average pooling.

The messaging passing mechanism in HG-Encoder is \emph{relation-wise}.
Representations of neighboring nodes connected to a user $u$ by the same relation $r$ are aggregated by three different pooling methods. Results are concatenated and passed to a single-layer feedforward neural network to generate relation-$r$-based representation $\mathbf{m}_{N_u^r}$ for user $u$:
\begin{equation}
	\label{eq:intra_rel}
	\begin{aligned}
		\mathbf{x}_{N_u^r}^{(k+1)}&=\text{mean}(\mathbf{h}_{r,j_1}^{(k)},\dots,\mathbf{h}_{r,j_*}^{(k)})\hskip 1.5em \mathbf{h}_{N_u^r}^{(k+1)}=\mathbf{x}_{N_u^r}^{(k+1)}\oplus \mathbf{y}_{N_u^r}^{(k+1)}\oplus \mathbf{z}_{N_u^r}^{(k+1)}\\
		\mathbf{y}_{N_u^r}^{(k+1)}&=\text{max}(\mathbf{h}_{r,j_1}^{(k)},\dots,\mathbf{h}_{r,j_*}^{(k)})\hskip 2em \mathbf{m}_{N_u^r}^{(k+1)}=\mathbf{W}_{r,m}\mathbf{h}_{N_u^r}^{(k+1)}+\mathbf{b}_{r,m}\\
		\mathbf{z}_{N_u^r}^{(k+1)}&=\text{sum}(\mathbf{h}_{r,j_1}^{(k)},\dots,\mathbf{h}_{r,j_*}^{(k)})
	\end{aligned}
\end{equation}
where the superscript $(k)$ indicates the $k$-th iteration, $\oplus$ is the concatenation operation, $N_u^r$ denotes the relation-$r$-based neighbors of node $u$ and $j_*\in N_u^r$. $\mathbf{h}_{r,j_*}^{(k)}$ is the representation of node $j_*$ for relation $r$, and $\mathbf{h}_{r,j_*}^{(0)}$ is equivalent to $\mathbf{h}_{j_*}^{(0)}$. $\text{max}(\cdot)$ and $\text{sum}(\cdot)$ are max pooling and sum pooling, respectively. $\mathbf{W}_{r,m}$ and $\mathbf{b}_{r,m}$ are parameters for the relation $r$. 

HG-Encoder adds a self-connection to each user node to retain original user features during message passing:
\begin{equation}
	\label{eq:self_con}
	\mathbf{s}_{r,u}^{(k+1)}=\mathbf{W}_{r,s}\mathbf{h}_{r,u}^{(0)}+\mathbf{b}_{r,s},\hskip 1em \mathbf{e}_{r,u}^{(k+1)}=\text{RELU}(\mathbf{m}_{N_u^r}^{(k+1)}\oplus \mathbf{s}_{r,u}^{(k+1)})
\end{equation}
where $\mathbf{W}_{r,s}$ and $\mathbf{b}_{r,s}$ are learnable parameters and $\text{RELU}(\cdot)$ is the Rectified Linear Unit. The acquired $\mathbf{e}_{r,u}$ is passed to a feedforward neural network with an $L_2$ normalization:
\begin{equation}
	\label{eq:l2}
	\mathbf{q}_{r,u}^{(k+1)}=\text{RELU}(\mathbf{W}_{r,q}\,\mathbf{e}_{r,u}^{(k+1)}+\mathbf{b}_{r,q}),\hskip 1em \mathbf{h}_{r,u}^{(k+1)}=\mathbf{q}_{r,u}^{(k+1)}\Big/\Big\lVert \mathbf{q}_{r,u}^{(k+1)}\Big\rVert
\end{equation}
where $\mathbf{W}_{r,q}$ and $\mathbf{b}_{r,q}$ are learnable weights. 

With $R$ relations, we can obtain $R$ relation-specific embeddings $\{\mathbf{h}_{1,i}^{(k+1)},\dots,\mathbf{h}_{R,i}^{(k+1)} \}$ for a user node $u$. Since one relation-specific node embedding only represents the node from one perspective, 
we further design an inter-relation aggregation module to learn more comprehensive user  representations. The inter-relation aggregation module uses a relation-level attention mechanism to fuse different relation-wise representations of a user $u$ into its overall representation $\mathbf{h}_u^{(k+1)}$:
\begin{equation}
	\label{eq:inter-rel}
	\begin{aligned}
	\mathbf{h}_u^{(k+1)}&=\sum_{r=1}^{R}\beta_r\cdot \mathbf{h}_{r,u}^{(k+1)},\hskip 1em \beta_r=\frac{\exp(w_r)}{\sum_{i=1}^R\exp(w_i)}\\
	w_r&=\frac{1}{|V|}\sum_{v\in V}\mathbf{a}^T\cdot\tanh(\mathbf{W}_{r,w}\cdot \mathbf{h}_{r,v}^{(k+1)}+\mathbf{b}_{r,w})
	\end{aligned}
\end{equation}
where $\beta_r$ weighs the importance of relation $r$, $\mathbf{W}_{r,w}$ and $\mathbf{b}_{r,w}$ are learnable weights, and $\mathbf{a}$ denotes the relation-level attention vector (i.e., a learnable parameter vector).

HG-Encoder stacks two of the above GNN layers (Eqs.~\ref{eq:intra_rel},~\ref{eq:self_con},~\ref{eq:l2} and~\ref{eq:inter-rel}) to generate the final representation $\mathbf{h}_{u}^{(k+1)}$ for user $u$. 
Some user nodes only exist starting from a certain snapshot of HTG. For other snapshots where they are absent, corresponding passing messages will be ignored.  
Note that, due to the large volume of MMMA data, we only apply the messaging passing mechanism to user nodes to reduce the training cost. 
Group/device representations are the average of their neighboring users' representations.

HG-Encoder can be connected to a score module defined as a linear mapping layer followed by a sigmoid function to estimate the abnormality of a user node.
Then, HG-Encoder can be optimized with a standard binary cross entropy loss over labeled user nodes. 
Note that HG-Encoder does not maintain node embeddings binding to specific nodes. 
Instead, learnable transformation weights $\mathbf{W}$, $\mathbf{b}$ and $\mathbf{a}$ are updated during optimization.
Hence, it is \emph{inductive} and can generate representations for new nodes that emerge every day in MMMAs. 

\subsection{Graph Dynamics Encoding}
\label{sec:pretrain}

To capture the dynamics of HTG and alleviate the reliance on supervision, we design special dynamics encoding method for \model and it mainly includes three parts: (1) constructing user history sequences, (2) augmenting user history sequences and (3) contrastive multi-view sequence encoding.

\begin{figure}[t]
	\begin{center}
		\includegraphics[width=1\linewidth]{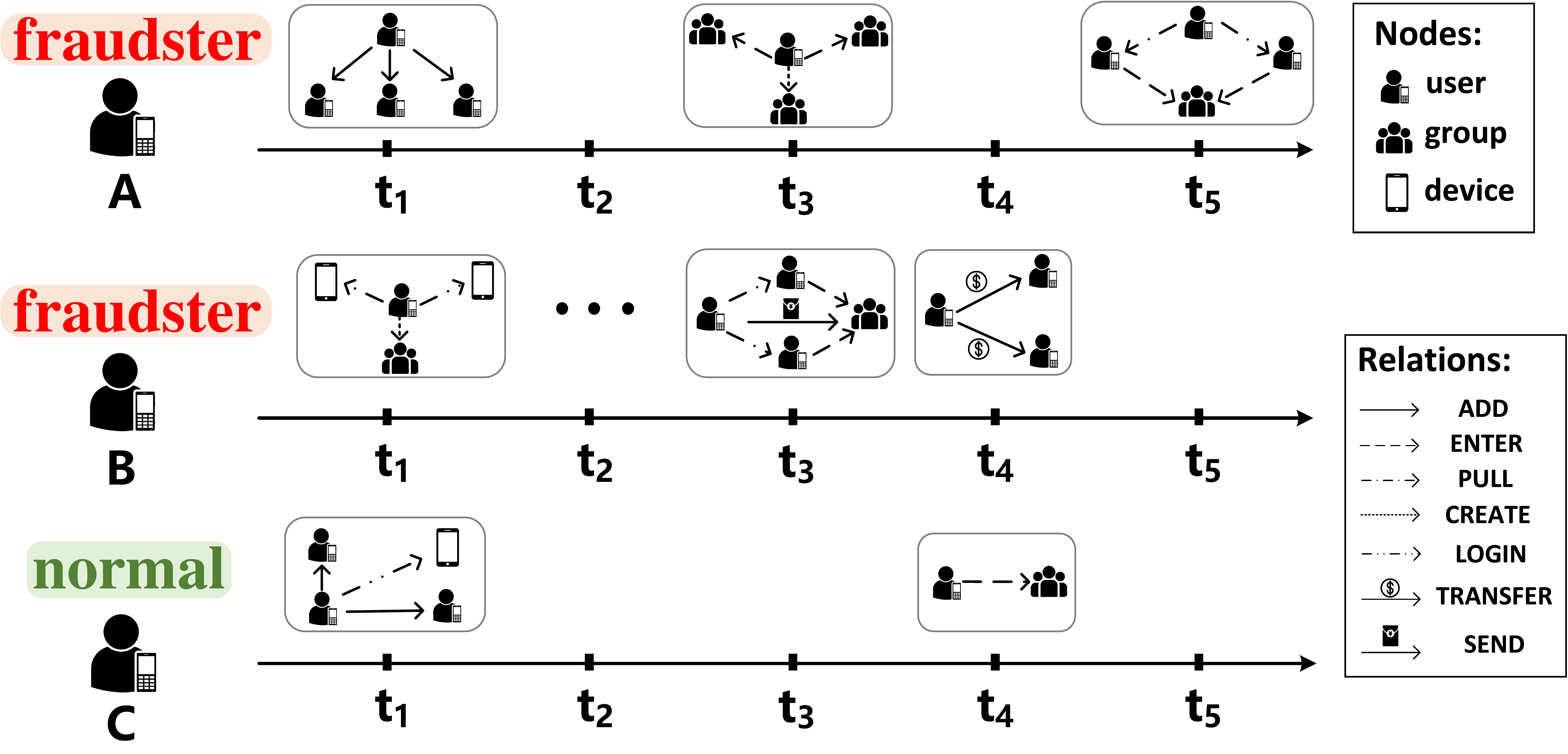}
	\end{center}
	\caption{Different users' behavioral sequences.}
	\label{fig:fraud_seq}
\end{figure}

\subsubsection{User History Sequences}
\label{sec:user_seq}

\model extracts and models two types of user history sequences from HTG, namely \emph{temporal snapshot sequence} and \emph{user relation sequence} to capture graph dynamics. 
They provide two different \emph{views} of user historical data.

\vspace{5pt}
\noindent\textbf{Temporal Snapshot Sequence.}
To perpetrate crowdsourcing fraud, the crucial part is gaining victims' trust. 
Fraudsters take a series of actions to achieve this goal. 
In our observation, \emph{a fraudster's diverse fraud actions may spread across several time points}. 
As depicted in Fig.~\ref{fig:fraud_seq}. 
A fraudster $A$ first \ADD many users, and spends a few days to use high rewards as bait to \PULL potential victims to \ENTER the fraud group, which is early \CREATE by the fraudster. 
Another fraudster $B$ first \LOGIN on many devices within a short time period, and then follow the same actions of fraudster $A$ ($t_1$ and $t_3$ in $A$'s timeline) to search for potential victims. 
Differently, after that, $B$ \SEND bonus packets as an encouragement and reassures victims that the reward is real. 
Later, $B$ \TRANSFER money with many users to either pay the reward to victims as bait or transfer the obtained, illicit money to accomplices. 
On the contrary, a normal user $C$ typically does not have so many behaviors within a relatively short period.

Based on our observation that the diverse fraud actions spread across different time points, the evolving characteristic of HTG can assist in distinguishing normal users and fraudsters. 
Therefore, we extract representations of a user in different snapshots as his/her temporal snapshot sequence to capture dynamics in the HTG. 
Given a series of historical snapshots $\{\mathcal{G}^t\}_{t=1}^{T}$, we apply HG-Encoder$_{\text{TSS}}$\footnote{TSS is short for temporal snapshot sequence.} over each snapshot to obtain the representations for all the nodes in each snapshot. HG-Encoder$_{\text{TSS}}$ adopts the encoder design illustrated in Sec.~\ref{sec:graph}. 
For each user node $u$, we define its temporal snapshot sequence as $\text{seq}^{\text{temp}}_{u}=[\mathbf{h}_u^1,\mathbf{h}_u^2,\dots, \mathbf{h}_u^{T}]$.

\vspace{5pt}
\noindent\textbf{User Relation Sequence.}
In HTG, a user's \emph{direct} actions that manifest in edges between itself and $1$-hop out-neighbors reveal his/her characteristics. 
We can observe from Figs.~\ref{fig:illustration} and~\ref{fig:fraud_seq} that crowdsourcing frauds typically involve several direct actions (e.g., \ADD, \PULL and \POST) of fraudsters appearing in different stages of frauds (i.e., search, gain trust and deceive).
Therefore, it is beneficial to model user relation sequence composed of edges in all the $1$-hop subgraphs of a user node from different snapshots.

Given a user node $u$ and its $1$-hop out-neighbors node set $N_u=[n_{u,1},\dots,n_{u,m}]$ associated with the edge set $E_u=\{(u,n_{u,1},t_1),\dots,(u,n_{u,m},t_m)\}$ where $m$ is the number of out-neighbors of $u$ and a tuple $(u,n_{u,i},t_i)$ indicates that there is an edge from $u$ to $n_{u,i}$ at time point $t_i$. 
$n_{u,i}$ can be a user node, a group node or a device node. 
If multiple possible nodes exist for $n_{u,i}$ at $t_i$, we sample one or more of them as $n_{u,i}$ as long as the number of the extracted user relation sequences is less than a threshold. 
For any two nodes $n_{u,i}$ and $n_{u,j}$ in the constructed sequence with $i<j$, their associated edges $(u,n_{u,i},t_i)$ and $(u,n_{u,j},t_j)$ satisfy $t_i<t_j$. 
Some examples are provided in the upper right corner of Fig.~\ref{fig:overview}.
\emph{User relation sequences describe user's behaviors over time, which remedies the limitation of temporal snapshot sequences that solely contain hidden states of the same user over time}. 

We use $1$-hop neighbors as they directly reveal a user's relations with other nodes, being a promising evidence to model a user's behavior pattern. 
Nevertheless, some issues require extra attention: 
\begin{itemize}
	\item The number of $1$-hop neighbors is uneven across the graph, meaning that some users are relatively active in the recorded time period while some are not. Active users have much more user relation sequences than inactive users, which may lead to model bias. Therefore, we sample up to a predefined maximum number of sequences for each user to avoid model bias as well as speed up model training. 
	\item We use directed edges to construct user relation sequences to maintain the asymmetry relation. 
	This design choice is based on our observation: fraudsters are usually more active than passive. They keep adding new friends to search for potential victims or pulling users into a chat group to commit fraud. 
	Each node in a directed graph plays two roles simultaneously: the source role (the source node of the directed edge) and the target role (the target node of the directed edge). Distinguishing the duality of nodes can often reveal some critical information detecting crowdsourcing fraud. 
	
	\item For users having few or no out-neighbors within the recorded time period, we take the sub-sequence from other users to construct user relation sequences. For instance, node $v_3$ in the upper right corner of Fig. 2 in our submission has no out-neighbors. But $v_3$ is an out-neighbor of $v_1$. Hence, we extract the sub-sequence $v_3\rightarrow v_2\rightarrow v_5$, which starts with $v_3$, from the user relation sequence of $v_1$ as the user relation sequence for $v_3$. 
	
\end{itemize}

\subsubsection{Data Augmentation.} 
\label{sec:data_aug}

\begin{figure}[t]
	\begin{center}
		\includegraphics[width=1\linewidth]{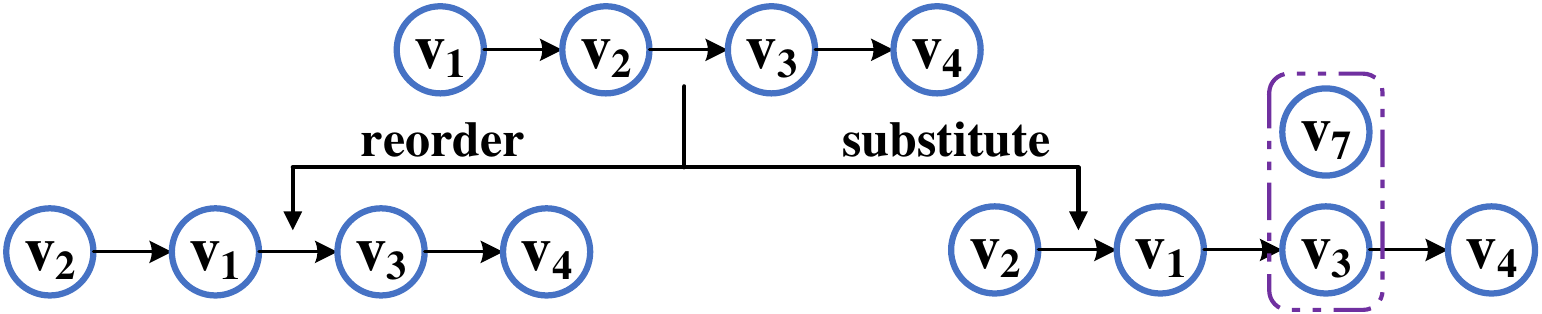}
	\end{center}
	\caption{Data augmentation in \model.}
	\label{fig:data_augment}
\end{figure}

We employ data augmentation to generate more meaningful sequences from temporal snapshot sequences and user relation sequences.
Specially, we design two augmentation approaches: \emph{reorder} and \emph{substitute}. 
As depicted in Fig.~\ref{fig:data_augment}, they can augment the same sequence from different views while preserving information of the original sequence. 
For each temporal snapshot sequence, we perform reordering twice to generate two augmented sequences.
For each user relation sequence, we also generate two augmented sequences. 
But one is from reordering and the other is from substitution.

\begin{figure}[t]
	\begin{center}
		\includegraphics[width=1\linewidth]{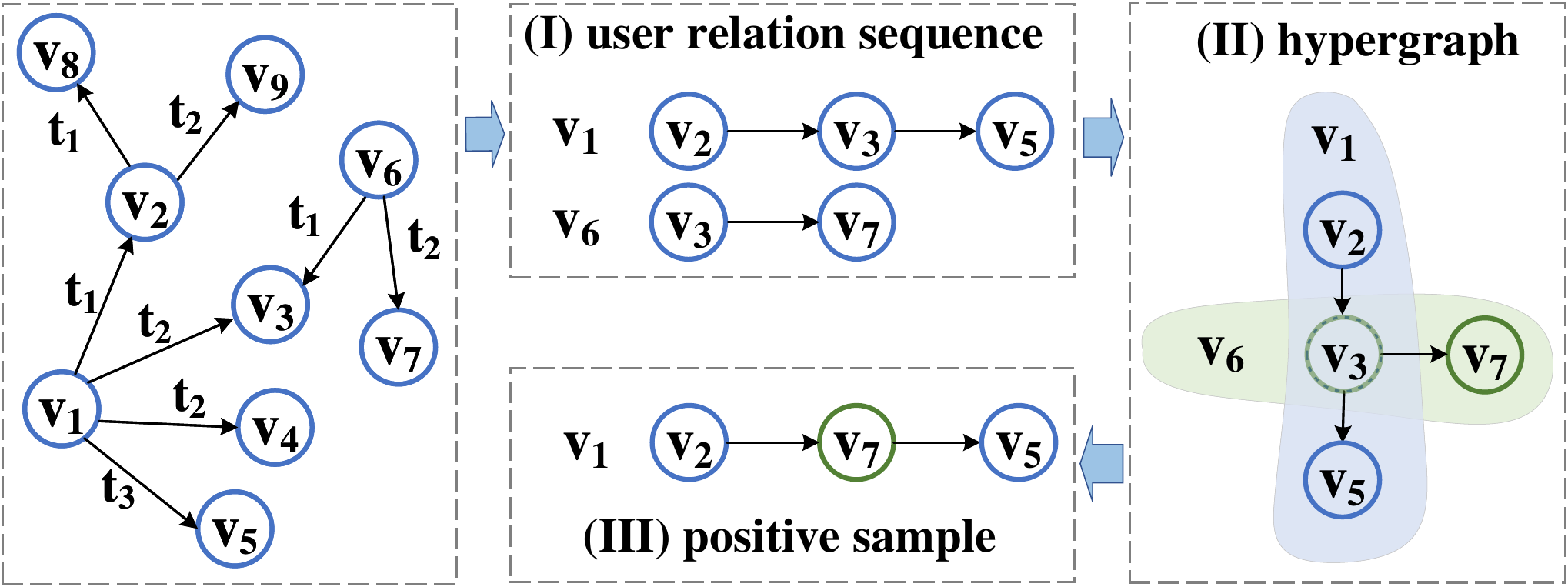}
	\end{center}
	\caption{An example for substitution.}
	\label{fig:substitution}
\end{figure}

\vspace{5pt}
\noindent\textbf{Reorder.} 
Fraudsters usually perpetrate crowdsourcing frauds as a gang rather than an individual cybercriminal. A fraudster may \PULL victims into a group created by a fraud conspirator, and then \CREATE another group for future deceptions. 
Hence, the two behavior order ``\CREATE-\PULL'' and ``\PULL-\CREATE'' are both meaningful in detecting frauds. 
Based on the above observation, we adopt the reordering operator to produce augmented sequences. 
Formally, given a reordering ratio $\gamma$ and a user history sequence $s_u$, we randomly shuffle a continuous subsequence in $s_u$ with length $\lfloor\gamma*|s_u| \rfloor$ to generate another sequence $s'_u$. $|s_u|$ indicates the length of $s_u$ and $s'_u$ has the same length (i.e., $|s_u|$) as $s_u$. 

\begin{table*}[ht]
  \begin{minipage}[b]{0.58\linewidth}
    \centering
    \includegraphics[width=\linewidth]{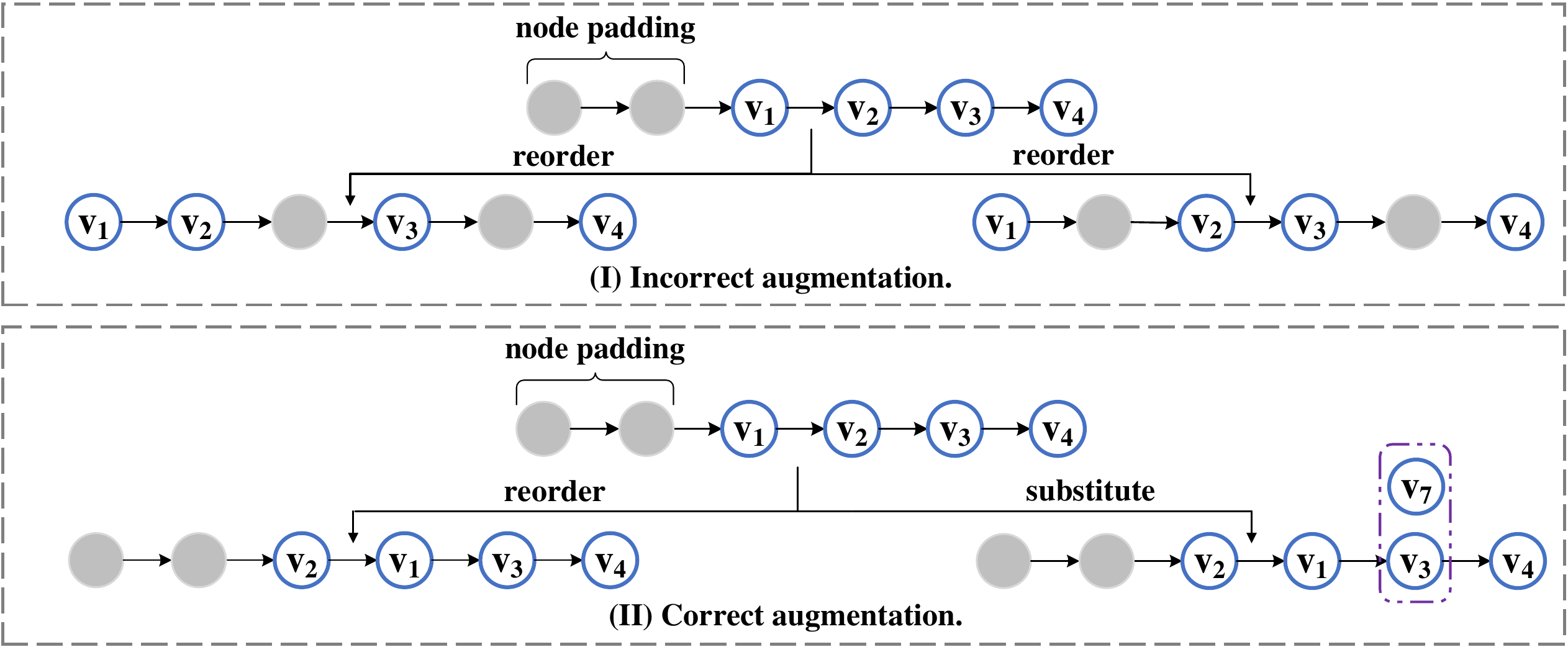}
    \captionof{figure}{Node padding in data augmentation.}
    \label{fig:incorrect_reorder}
  \end{minipage}
  \hfill
  \begin{minipage}[b]{0.38\linewidth}
    \centering
    \includegraphics[width=\linewidth]{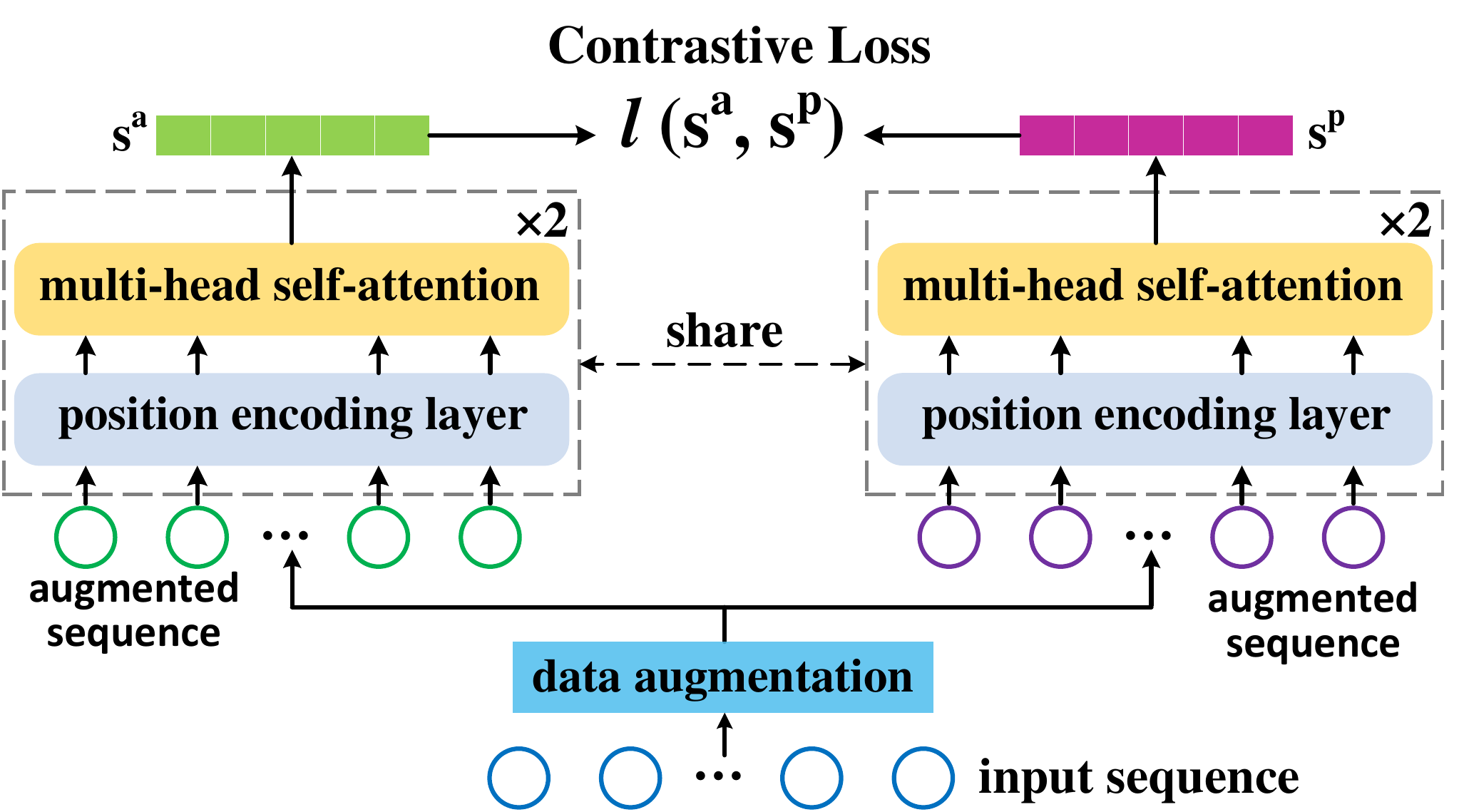}
    \captionof{figure}{Overview of CS-Encoder.}
    \label{fig:seq_encoder}
  \end{minipage}
\end{table*}

\vspace{5pt}
\noindent\textbf{Substitute.} The main idea of the substitution operator is to find a similar element to the selected one in the input user relation sequence. The key is to choose the appropriate, new element that roughly preserves the same information as the selected, old one for substitution. 
We propose to utilize the \emph{hypergraph} to pick such similar elements. 
As shown in Fig.~\ref{fig:substitution}, one input user relation sequence $v_1$: $v_2\rightarrow v_3\rightarrow v_5$ shares $v_3$ with another user relation sequence $v_6$: $v_3\rightarrow v_7$. These two sequences can naturally form a hypergraph with two hyperedges denoted by colored areas in Fig.~\ref{fig:substitution}(II). 
Hyperedges help reveal the hidden connection unavailable in the original HTG. 
For example, $v_1$ and $v_6$ are both fraudsters, $v_3$ is a normal group, and $v_7$ is actually a fraud group. 
In the original HTG, $v_1$ has no direct connection with $v_7$, which may be the camouflage for escaping detection. 
A fraudster can leverage multiple \app accounts to avoid joining too many groups and revealing frauds. 
With the help of the hypergraph, we can find that $v_1$ and $v_7$ has a hyperconnection via $v_3$, which reveals the hidden information.
For this case, $v_3$ is an intermediate group that may be the camouflage, while $v_7$ is the true ending node (the fraud group) that plays an pivotal role in the fraud.
Thus, we can replace $v_3$ with $v_7$ to generate another user relation sequence $v_2\rightarrow v_7\rightarrow v_5$ that can be used as a positive view of the original user relation sequence of $v_1$. 
More specially, given a substitution ratio $\delta$ and a user relation sequence $s_u^{\text{usr}}=[s_u^1,s_u^2,\dots,s_u^{|s_u|}]$, we randomly select $\lfloor \alpha*|s_u| \rfloor$ nodes.
For each element $s_u^i$ in the selected element set and its associate hyperedge set $E_i^{\text{hyper}}$, we randomly select a hyperedge $e$ from $E_i^{\text{hyper}}$ and use the element $s_u^j$ connected to $s_u^i$ via $e$ to substitute $s_u^i$.

\vspace{5pt}
\noindent\textbf{Implementation Issue of Data Augmentation.} 
In augmentation, the two types of user history sequences may contain missing nodes at its beginning or its end, which requires padding. The reason is that a user may only exist in part of the snapshots of HTG or have fewer actions compared to other users. 
Directly augmenting on the complete sequence may generate \emph{broken} sequences. 
As shown in Fig.~\ref{fig:incorrect_reorder}(I), reordered sequences are broken by padding nodes. 
Therefore, we only perform augmentation on the non-padding part of the user history sequence, as shown in Fig.~\ref{fig:incorrect_reorder}(II).

\subsubsection{Contrastive Sequence Encoder.}
\label{sec:seq_encoder}

As shown in Fig.~\ref{fig:seq_encoder}, we design a \underline{C}ontrastive \underline{S}equence encoder (CS-Encoder) equipped with Transformer and contrastive learning~\cite{abs-2105-07342} to encode user history sequences into informative representations. 

We adopt a position encoding matrix $\mathbf{P}\in \mathbb{R}^{T\times d}$ to preserve the order of a sequence, where $T$ is the time span of the HTG. 
Given an input sequence representation $s=[\mathbf{h}_1,\mathbf{h}_2,\dots,\mathbf{h}_T]$, the position embeddings are injected to construct the position-aware input to the multi-head self-attention module with eight heads: $\mathbf{h}_t^0=\mathbf{h}_t+\mathbf{p}_t$, 
where $\mathbf{p}_t$ is the position encoding of the time point $t$ and $1\le t \le T$. 
$\mathbf{h}_t^0$ is then fed into the multi-head self-attention module in Transformer to generate the sequence representation. 
The position encoding layer and the multi-head self-attention module constitute an encoding block. 
We stack two encoding blocks in CS-Encoder.
We use the last hidden state vector in the output from the second block as the sequence representation.

Recall that we design two types of user history sequences and we have two encoders in \model to encode each of them:
\begin{itemize}
	\item As shown in top left corner of Fig.~\ref{fig:overview}, \model uses the TSS-Encoder for encoding temporal snapshot sequences. TSS-Encoder includes an HG-Encoder$_{\text{TSS}}$ and a CS-Encoder. HG-Encoder$_{\text{TSS}}$ encodes the temporal snapshot sequence of a user $u$ to $\text{seq}^{\text{temp}}_{u}$, which is later fed to the CS-Encoder. CS-Encoder generates the sequence representation of the temporal snapshot sequence. 
	
	\item As shown in top right corner of Fig.~\ref{fig:overview}, \model uses the URS-Encoder\footnote{URS is short for user relation sequence.} for encoding user relation sequences. URS-Encoder adopts the same projection method as Eq.~\ref{eq:project} to generate node representations for constructingx the input sequence representation to the CS-Encoder. Then, the CS-Encoder in the URS-Encoder generates the representation of user relation sequence.
\end{itemize}

\vspace{5pt}
We adopt contrastive learning that learns by contrasting between similar and dissimilar objects instead of recognizing them one by one to help TSS-Encoder and URS-Encoder capture the differences between fraudsters and normal users.
Recall that, through data augmentation, each user history sequence has two additional augmented sequences. 
TSS-Encoder and URS-Encoder learn to maximize the differences between augmented sequences derived from different users and minimize the differences between the two augmented sequences generated from the same user.

Given a mini-batch of $N$ users and one type of their corresponding user history sequence, we have augmented user history sequences of size $2N$. 
For any user $u$, we treat the representation $\mathbf{s}_{u}^a$ of its one augmented sequence as the anchor, and the representation $\mathbf{s}_{u}^p$ of the other augmented sequence naturally becomes the positive sample.
The rest $2(N-1)$ augmented sequences within the same batch are treated as negative samples.
The contrastive loss function for each positive pair $\langle \mathbf{s}_u^a, \mathbf{s}_u^p\rangle$ is shown as follows:
\begin{equation}
	\label{eq:cl_loss}
	l(\mathbf{s}_u^a, \mathbf{s}_u^p)=\log\frac{e^{\theta(\mathbf{s}_u^a, \mathbf{s}_u^p)/\tau}}{e^{\theta(\mathbf{s}_u^a, \mathbf{s}_u^p)/\tau}+\sum_{v=1}^N \mathbbm{1}_{[v\not =u]} (e^{\theta(\mathbf{s}_u^a, \mathbf{s}_v^a)/\tau}+e^{\theta(\mathbf{s}_u^a, \mathbf{s}_v^p)/\tau})}
\end{equation}
where $\mathbbm{1}_{[v\not =u]}\in \{0,1\}$ is the indicator function that equals 1 if $v\not =u$ otherwise 0, and $\tau$ is a temperature parameter. 
We adopt the cosine similarity to calculate $\theta(u,v)=\text{sim}\big(g(u), g(v)\big)$ where $g(\cdot)$ is a two-layer MLP. 

Since we can exchange two views of a user history sequence (i.e., the two augmented sequences) in the loss function defined in Eq.~\ref{eq:cl_loss}, we can similarly define another loss by using $\mathbf{s}_u^p$ as anchor. 
The contrastive learning loss used to optimize TSS-Encoder/URS-Encoder is defined as the average over all positive pairs: 
\begin{equation}
\mathcal{L}_{\text{cl}}=\frac{1}{2N}\sum_{i=1}^N[l(\mathbf{s}_u^a, \mathbf{s}_u^p)+l(\mathbf{s}_u^p, \mathbf{s}_u^a)].
\end{equation}

We train both TSS-Encoder and URS-Encoder using a binary classification task (i.e., estimate whether the source of a sequence is a fraudster) together with the contrast task. 
For the binary classification task, TSS-Encoder/URS-Encoder is connected to a score module defined as a linear mapping layer followed by a sigmoid function, which is the same score module connected to HG-Encoder as described in Sec.~\ref{sec:graph}.
The suspicious score is passed to the binary cross-entropy loss $\mathcal{L}_{\text{binary}}$ which is calculated over limited labeled data.
The overall loss for optimizing TSS-Encoder/URS-Encoder is defined as a multi-tasking loss:  
\begin{equation}
	\label{eq:losses}
	\mathcal{L}=\mathcal{L}_{\text{binary}} + \mathcal{L}_{\text{cl}}.
\end{equation}

\subsection{Putting All Together}

In the pretraining phase, TSS-Encoder, URS-Encoder and the detection module $\text{HG-Encoder}_{\text{detect}}$ are trained independently on the training data with limited labels.  
$\text{HG-Encoder}_{\text{detect}}$ adopts the same design illustrated in Sec.~\ref{sec:graph} followed by the same score module as used in the binary classification task of TSS-Encoder/URS-Encoder.

During the detection phase, for a user $u$, we generated temporal snapshot sequence representation $\mathbf{h}_{\text{seq}^{\text{temp}}_{u}}$ and user relation sequence representation $\mathbf{h}_{\text{seq}^{\text{rel}}_{u}}$ using pretrained TSS-Encoder and URS-Encoder, respectively. 
Then, as shown in Fig.~\ref{fig:overview}(b), the two generated representations and the initial sequence representation $\mathbf{h}_u^{(0)}$ are concatenated and fed into HG-Encoder$_{\text{detect}}$ followed by its scoring module to estimate the suspicious score of $u$. If the estimate score for $u$ is larger than 0.5, $u$ will be predicted as a fraudster.

\vspace{5pt}
\noindent\textbf{Model Size.} The model size of \model is in the magnitude of $f^2$ where $f$ is the dimensionality of representations. Note that prevalent GNN-based methods involve at least one weight matrix and their model size is at least $f^2$. Hence, the model size of \model is in the same magnitude as other GNN-based methods. We provide the detailed analysis in Appendix~\ref{app:modelsize}.

\vspace{5pt}
\noindent\textbf{Deployment on Large Graphs.} As \model is designed for large graphs, we use a simple subgraph design instead of other sophisticated subgraph designs so that \model can conduct sampling before training to avoid high CPU-GPU I/O cost. 
Data augmentation is also performed before training to avoid high overhead.

\section{Experiments}
\label{sec:exp}

\subsection{Experiment Setting}

\noindent\textbf{Data.} 
We use two industry-size datasets for our experiments.
\begin{itemize}

\item \textbf{\app Dataset:} We receive the permission from \app for using their data. 
We construct a large-scale HTG as defined in Sec.~\ref{sec:graph_construct}. 
The graph contains nearly 6.8 million user nodes, 151 thousand \group nodes and 126 thousand device nodes. 
The number of edges are approximately 29.7 million covering 7 relations introduced in Sec.~\ref{sec:graph_construct}.
Only 53,660 user nodes are manually labeled by human experts: 10,749 of them are fraud users and 42,911 are normal users. 
The labels for other user nodes are unknown. 
We use one day as the time interval between two time point.
Thus, we derive 14 separate graph snapshots from the HTG. 
We set the maximum number of sampled user relation sequence for each user to be 10.
We randomly divide the labeled users by a ratio of 8:1:1 for training, validation and test. Unlabeled users are available during training, validation and test.

\item \textbf{FinGraph Dataset\footnote{\url{https://ai.ppdai.com/mirror/goToMirrorDetailSix?mirrorId=28}}:} This dataset is provided by the 7th Finvolution Data Science Competition for evaluating financial fraud detection. It contains an anonymized dynamic user-user interaction graph in financial industry. It has approximately 4.1 million user nodes and 5 million edges. Every node has 17 features. Edges have 11 types. There are 82 thousand labeled user nodes: 1 thousand nodes are financial fraudsters and 81 thousand nodes are normal users. 
The competition does not disclose node identity and physical meanings of node feature, node types and edge types for protecting privacy.
We randomly divide the labeled users by a ratio of 8:1:1 for training, validation and test. Unlabeled users are available during training, validation and test.

\end{itemize}

\vspace{5pt}
\noindent\textbf{Baselines.} \emph{We investigate existing GAD methods and find most of them cannot handle an industry-size MMMA graph. We choose the following baselines that can be used on \app and FinGraph:}
\begin{itemize}
	\item \textbf{Non-GNN classification methods}: XGBoost~\cite{ChenG16} and MLP. MLP is a feedfoward neural network with three hidden layers and it has 128, 64, 32 neurons in each layer, respectively.
	\item \textbf{Homogeneous graph based methods}: GCN~\cite{KipfW17} and GAT~\cite{VelickovicCCRLB18}.
	\item \textbf{Heterogeneous graph based methods}: HG-Encoder illustrated in Sec.~\ref{sec:graph}, RGCN~\cite{SchlichtkrullKB18} and Simple-HGN~\cite{LvDLCFHZJDT21}.
	\item \textbf{Graph based anomaly detection methods}: DCI~\cite{Wang0GYL021} and GeniePath~\cite{LiuCLZLSQ19}. DCI adopts contrastive learning for GAD.
	\item \textbf{Temporal GAD methods}: AddGraph~\cite{ZhengLLLG19}. Furthermore, we replace GCN in AddGraph with RGCN to model the heterogeneous information and name this variant as AddGraph-H.
\end{itemize}

We adopt the same score module design as \model for baselines without a score module. 
All methods adopt Adam optimizer with initial learning rate being 0.001 if possible. 
We use 64 as the dimension of representations and set the batch size to 256. 
The reordering and substitution ratios $\gamma$ and $\alpha$ are set to be 0.4.
All methods are terminated when they converge.

\vspace{5pt}
\noindent\textbf{Evaluation Metrics.} 
We use AUC and KS as metrics. 
AUC signifies the probability that the positive sample's score is higher than the negative sample's score. 
KS measures the degree of separation between the positive and negative distributions~\cite{2280095}.
KS is widely used in financial anomaly detection~\cite{AlarajAM21,li2020comparative}.

\subsection{Overall Detection Results.} Tab.~\ref{tab:overall_per} presents the overall detection results of all methods. 
We analyze the experimental results as follows:
\begin{enumerate}
	\item GNN-based approaches GCN and GAT generally exceed non-GNN methods XGBoost and MLP, indicating the spatial dependencies depicted by graph structure contain rich information that can improve detection performance.
	\item HG-Encoder significantly outperforms other GNN-based methods. This observation supports our decision of using HG-Encoder as the backbone of \model.
	\item Both dynamic and heterogeneous graph based models achieve satisfactory results, and heterogeneous graph based methods generally outperform homogeneous graph based approaches. This observation shows that temporal dependencies and multi-relation information help model user behavior pattern better and improve the detection.
	\item \model achieves much better performance than other baselines including state-of-the-art dynamic graph anomaly detection methods AddGraph and AddGraph-H. It consistently outperforms all baselines on both measures on both \app and FinGraph datasets. The results show that \emph{\model is superior to baselines on crowdsourcing fraud detection and can also be applied in other GAD tasks}.
\end{enumerate}

\begin{table}[t]
	\centering
	\caption{Overall detection performance. Results of \model and best baselines are shown in bold.}
	\label{tab:overall_per}
	\scalebox{0.85}{
		\begin{tabular}{c|cc|cc}
			\hline
			& \multicolumn{2}{c|}{\textbf{WeChat}}                   & \multicolumn{2}{c}{\textbf{FinGraph}}                  \\ \hline
			\textbf{Method}     & \multicolumn{1}{c|}{\textbf{AUC}}    & \textbf{KS}     & \multicolumn{1}{c|}{\textbf{AUC}}    & \textbf{KS}     \\ \hline
			XGBoost             & \multicolumn{1}{c|}{0.7189}          & 0.3281          & \multicolumn{1}{c|}{0.7388}          & 0.3911          \\ \hline
			MLP                 & \multicolumn{1}{c|}{0.7188}          & 0.3404          & \multicolumn{1}{c|}{0.7404}          & 0.4072          \\ \hline
			GCN                 & \multicolumn{1}{c|}{0.8082}          & 0.4790          & \multicolumn{1}{c|}{0.7530}          & 0.4205          \\ \hline
			GAT                 & \multicolumn{1}{c|}{0.7967}          & 0.4591          & \multicolumn{1}{c|}{0.7762}          & 0.4496          \\ \hline
			RGCN                & \multicolumn{1}{c|}{0.8400}          & 0.5361          & \multicolumn{1}{c|}{0.8002}          & 0.5217          \\ \hline
			Simple-HGN          & \multicolumn{1}{c|}{0.8484}          & 0.5474          & \multicolumn{1}{c|}{0.7798}          & 0.4669          \\ \hline
			GeniePath           & \multicolumn{1}{c|}{0.8234}          & 0.5257          & \multicolumn{1}{c|}{0.8115}          & 0.5466         \\ \hline
			DCI                 & \multicolumn{1}{c|}{0.8328}          & 0.5397          & \multicolumn{1}{c|}{0.7737}          & 0.4440          \\ \hline
			HG-Encoder          & \multicolumn{1}{c|}{\textbf{0.8682}}          & \textbf{0.5905}          & \multicolumn{1}{c|}{\textbf{0.8194}}          & \textbf{0.5485}          \\ \hline
			AddGraph            & \multicolumn{1}{c|}{0.8221}          & 0.4924          & \multicolumn{1}{c|}{0.7488}          & 0.3818          \\ \hline
			AddGraph-H          & \multicolumn{1}{c|}{0.8452}          & 0.5365          & \multicolumn{1}{c|}{0.8013}          & 0.5236          \\ \hline
			CMT                 & \multicolumn{1}{c|}{\textbf{0.9014}} & \textbf{0.6624} & \multicolumn{1}{c|}{\textbf{0.8354}} & \textbf{0.5720} \\ \hline
		\end{tabular}
	}
\end{table}

\begin{table}[!t]
	\centering
	\caption{Results of the ablation study. Best results are in bold.}
	\label{tab:3}
	\scalebox{0.85}{
		\begin{tabular}{c|cc|cc}
			\hline
			& \multicolumn{2}{c|}{\textbf{WeChat}}                   & \multicolumn{2}{c}{\textbf{FinGraph}}                  \\ \hline
			\textbf{Method}                                                 & \multicolumn{1}{c|}{\textbf{AUC}}    & \textbf{KS}     & \multicolumn{1}{c|}{\textbf{AUC}}    & \textbf{KS}     \\ \hline
			HG-Encoder                                                      & \multicolumn{1}{c|}{0.8682}          & 0.5905          & \multicolumn{1}{c|}{0.8194}          & 0.5485          \\ \hline
			$\text{\model}_{\text{TSS}}$                                    & \multicolumn{1}{c|}{0.8853}          & 0.6169          & \multicolumn{1}{c|}{0.8233}          & 0.5554          \\ \hline
			$\text{\model}_{\text{TSS}_{\text{cl}}}$                        & \multicolumn{1}{c|}{0.8957}          & 0.6461          & \multicolumn{1}{c|}{0.8331}          & 0.5683          \\ \hline
			$\text{\model}_{\text{URS}}$                                    & \multicolumn{1}{c|}{0.8911}          & 0.6337          & \multicolumn{1}{c|}{0.8267}          & 0.5554          \\ \hline
			$\text{\model}_{\text{URS}_{\text{cl}}}$                        & \multicolumn{1}{c|}{0.8929}          & 0.6347          & \multicolumn{1}{c|}{0.8298}          & 0.5625          \\ \hline
			$\text{\model}_{\text{TSS}-\text{URS}}$                         & \multicolumn{1}{c|}{0.8889}          & 0.6316          & \multicolumn{1}{c|}{0.8312}          & 0.5692          \\ \hline
			$\text{\model}_{\text{TSS}_{\text{cl}}-\text{URS}}$             & \multicolumn{1}{c|}{0.8999}          & 0.6624          & \multicolumn{1}{c|}{\textbf{0.8355}} & 0.5692          \\ \hline
			$\text{\model}_{\text{TSS}-\text{URS}_{\text{cl}}}$             & \multicolumn{1}{c|}{0.8946}          & 0.6413          & \multicolumn{1}{c|}{0.8324}          & 0.5703          \\ \hline
			$\text{\model}_{\text{TSS}_{\text{cl}}-\text{URS}_{\text{cl}}}$ & \multicolumn{1}{c|}{\textbf{0.9014}} & \textbf{0.6624} & \multicolumn{1}{c|}{0.8354}          & \textbf{0.5720} \\ \hline
		\end{tabular}
	}
\end{table}

\subsection{Ablation Study.} 

To verify the contribution of each component in \model, we report results of different variations of \model in Tab.~\ref{tab:3}:
\begin{itemize}
	\item \textbf{HG-Encoder}: it only uses Heterogeneous GNN encoder.
	\item \textbf{Variations with the subscript \text{TSS}}: it removes URS-Encoder and uses TSS-Encoder only.
	\item \textbf{Variations with the subscript \text{URS}}: it removes TSS-Encoder and uses URS-Encoder only.
	\item \textbf{Variations with the subscript \text{TSS-URS}}: both TSS-Encoder and URS-Encoder are used.
	\item \textbf{Smaller subscript ``cl'' for either TSS or URS}: the contrastive loss $\mathcal{L}_{\text{cl}}$ is used together with $\mathcal{L}_{\text{binary}}$ for optimizing TSS-Encoder or URS-Encoder. For the subscript TSS or URS without the smaller subscript ``cl'', only $\mathcal{L}_{\text{binary}}$ is used for optimizing TSS-Encoder or URS-Encoder. 
\end{itemize}

From the results in Tab.~\ref{tab:3}, we can observe that:
\begin{enumerate}
	\item From Tab.~\ref{tab:3}, we can see that most modules bring promising performance gain. The incorporation of either TSS Encoder or USR Encoder brings performance gain, as both \model-TSS and \model-USR significantly outperform HG-Encoder. In a few cases (e.g., compare CMT$_\text{TSS-URS}$ and CMT$_{\text{TSS}_{\text{cl}}-\text{URS}}$), some modules do not significantly improve the results on one dataset, but they show obvious enhancement on the other dataset.

	\item The complete \model (i.e., $\text{\model}_{\text{TSS}_{\text{cl}}-\text{URS}_{\text{cl}}}$) generally shows best results. For the case where it is not the best, the performance gap is subtle. Hence, modeling two views of historical data together can remedy the limitation of capturing only one view.
\end{enumerate}
Overall, we can conclude that each module in \model indeed contributes to the superior performance of \model.

\subsection{Sensitivity analysis of $\gamma$ and $\alpha$.} 

\begin{figure}[!t]
	\begin{center}
		\includegraphics[width=1\linewidth]{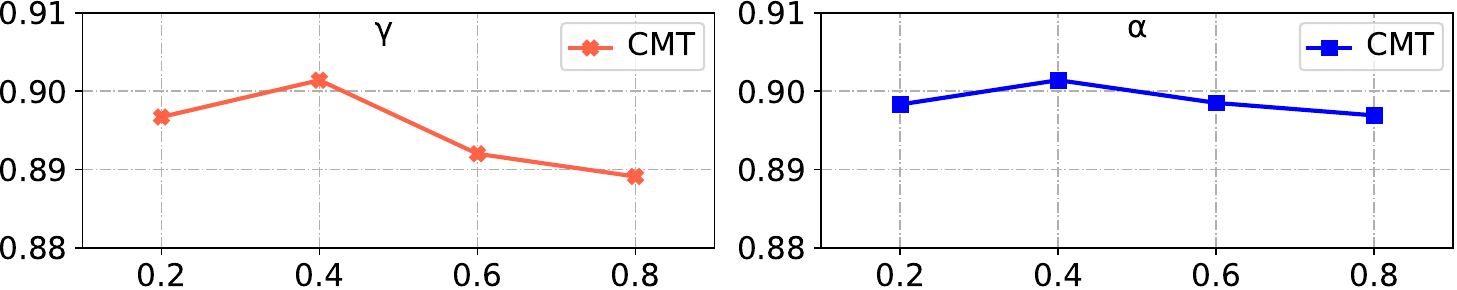}
	\end{center}
	\caption{Sensitivity analysis of $\gamma$ and $\alpha$ on \app dataset.}
	\label{fig:sen}
\end{figure}

Fig.~\ref{fig:sen} reports AUC of \model on \app dataset when using different values for $\gamma$ and $\alpha$.
We can observe that changes of $\gamma$ and $\alpha$ do not significantly affect the performance of \model. And the resulting AUC performance is around 0.89-0.9.
The observation shows that our data augmentation method is not sensitive to augmentation hyper-parameters.

\subsection{Quality of Representation.} 

\begin{figure}[!t]
	\centering
	\subfloat{{\includegraphics[width=0.45\linewidth]{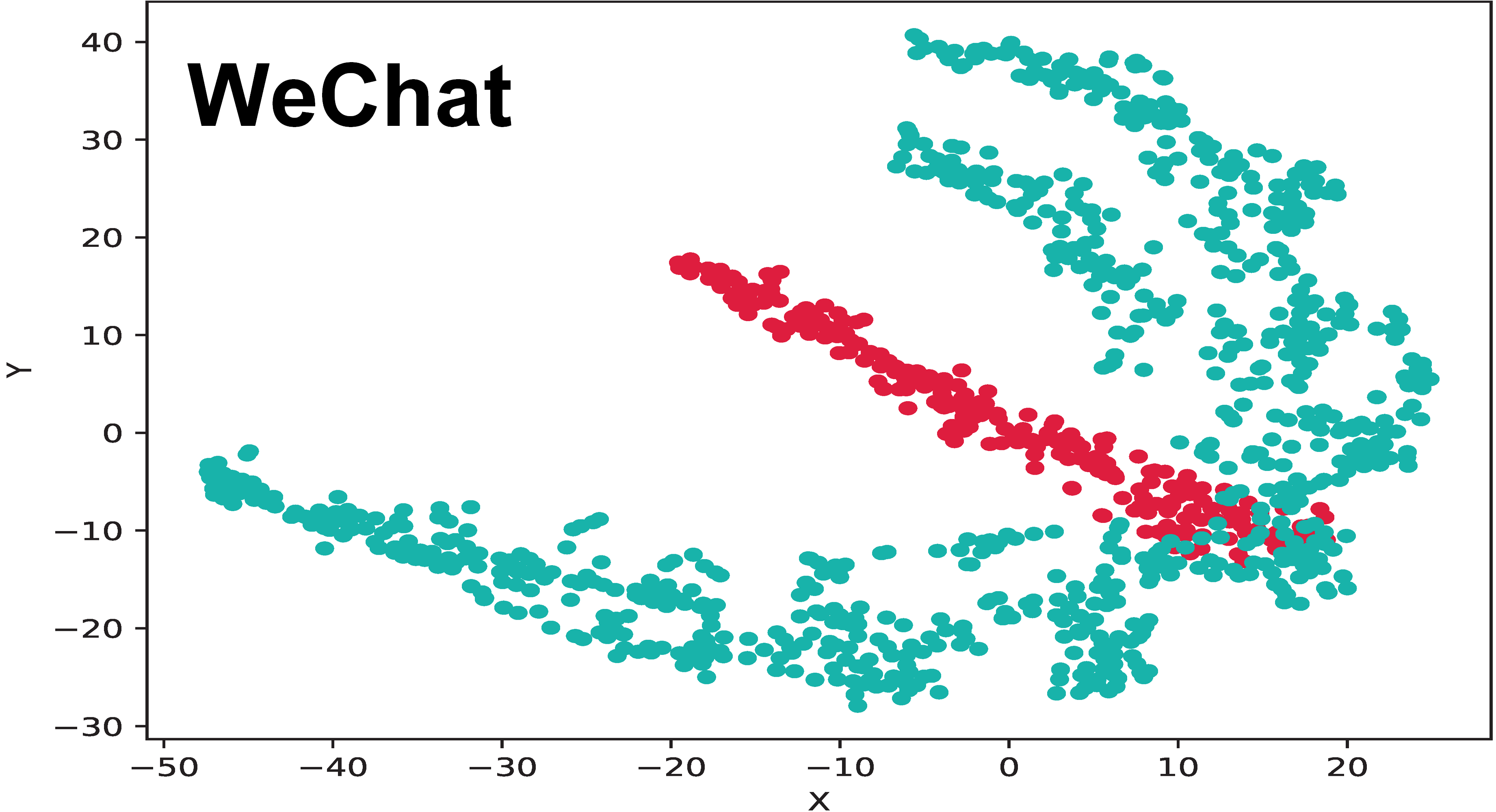} }}%
	\quad\quad
	\subfloat{{\includegraphics[width=0.45\linewidth]{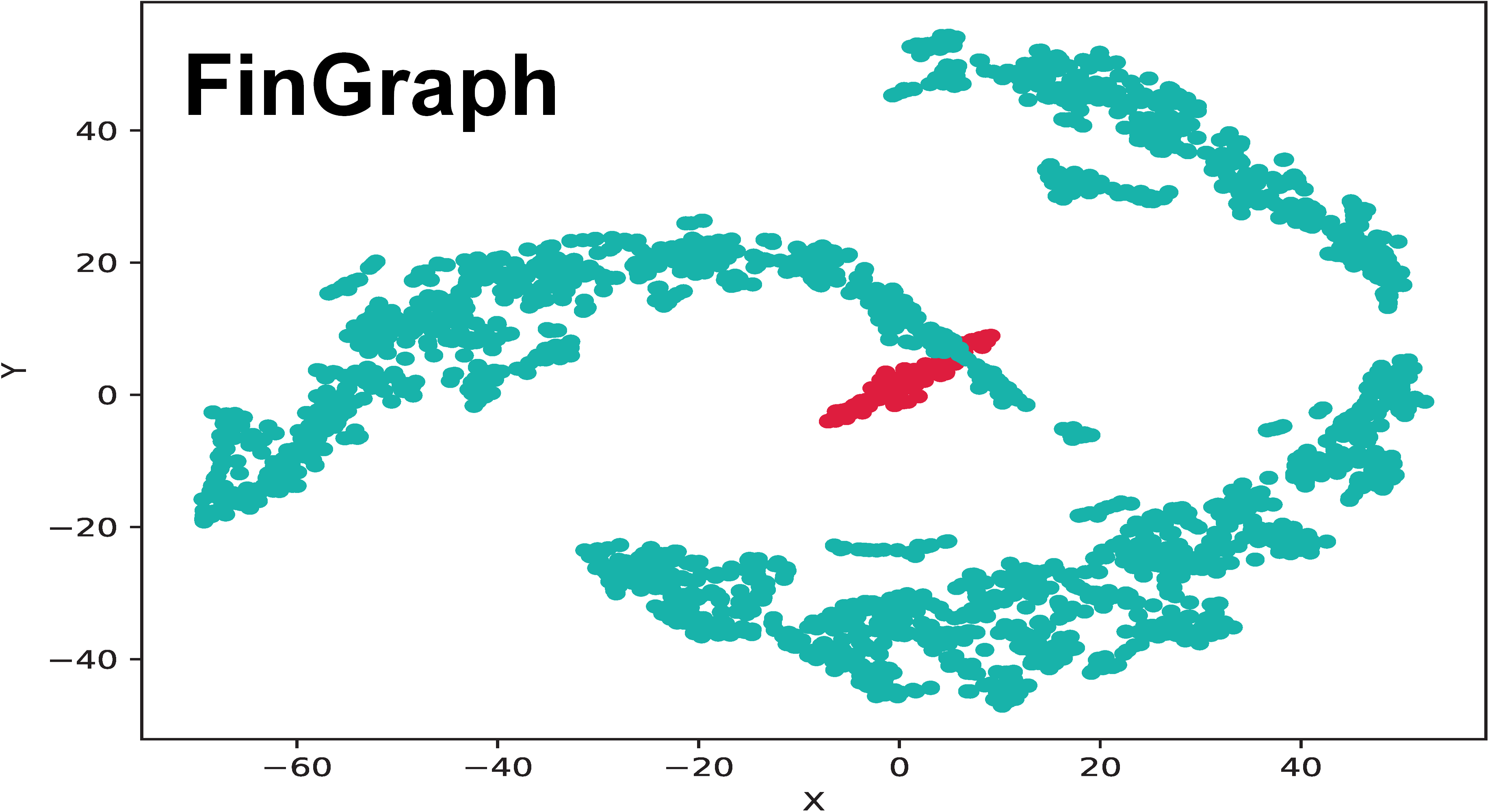} }}%
	\caption{t-SNE projection of user node representations generated by \model: (1) Red: fraudsters. (2) Green: normal users.}%
	\label{fig:vis}
\end{figure}

We adopt t-SNE~\cite{MaatenH08} to project representations of user nodes in the test set into a 2-dimensional space. From the result shown in Fig.~\ref{fig:vis}, we can see that representations of normal users and fraudsters have a clear distinction, showing that \model is able to produce high-quality representations for crowdsourcing fraud detection.

Moreover, the representations generated by the pretraining phase of \model can be utilized as additional input node features to other detection methods and improve the detection. We provide some discussions in Appendix~\ref{app:exp} of the supplementary material.

\subsection{Discovery of Fraud Patterns.} 

\begin{figure}[!t]
	\begin{center}
		\includegraphics[width=0.95\linewidth]{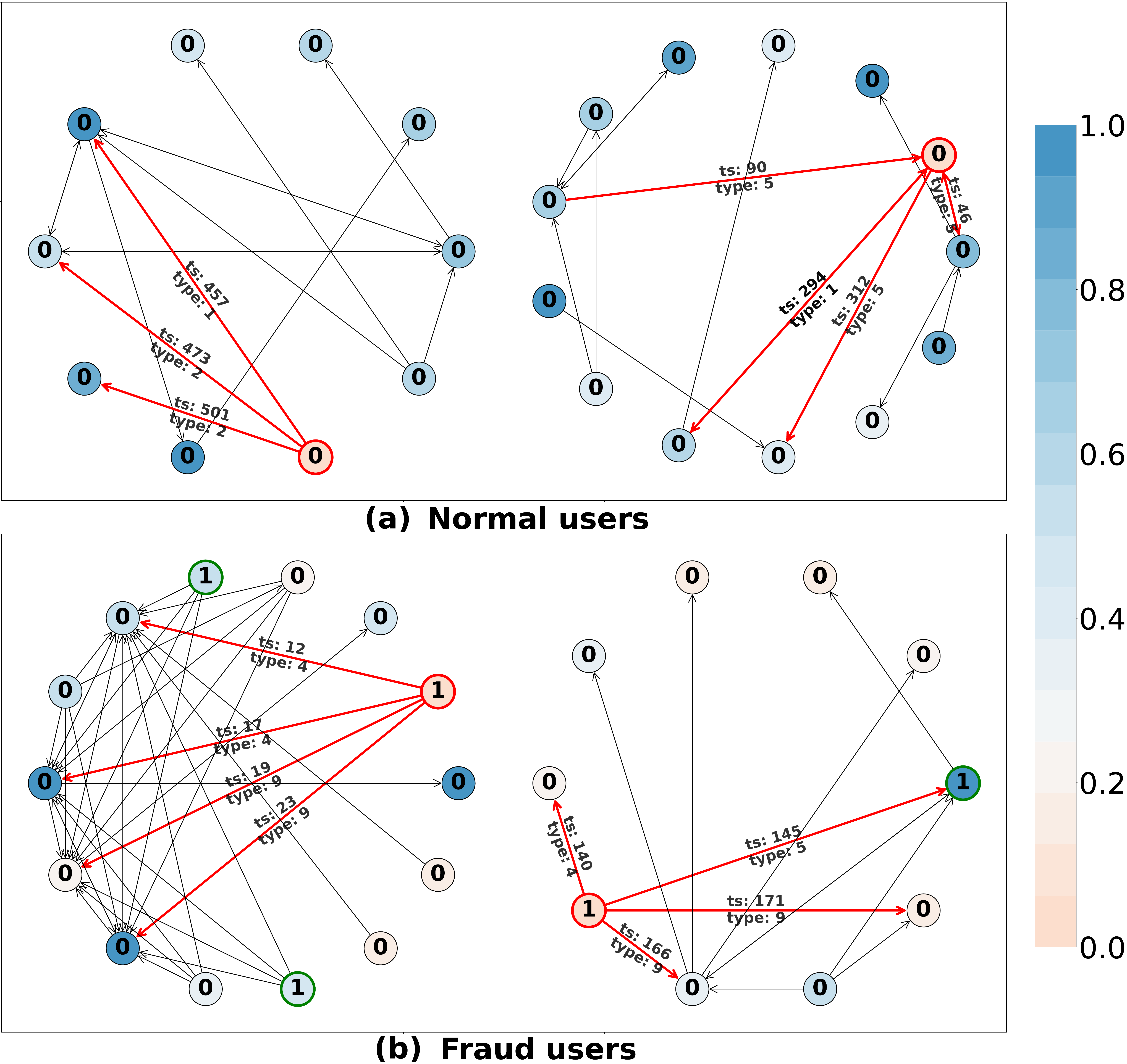}
	\end{center}
	\caption{Discovered fraud patterns on FinGraph dataset.}
	\label{fig:casestudy}
\end{figure}

We present some interesting fraud patterns discovered by using \model and they can help human investigators identify frauds.
Since we do not receive permission to disclose the patterns found in \app, we use the public FinGraph to illustrate.
Fig.~\ref{fig:casestudy} shows four subgraphs around two normal users and two fraudsters which are correctly identified by \model.
The four identified nodes are in red circles. 
Nodes label 1 indicates fraudsters while label 0 indicates normal users.
``TS'' over edges indicate the time when the corresponding interaction (edge) is observed.
The smaller the TS value is, the earlier the interaction appears.
We calculate the reciprocal Euclidean distance between representations of the identified node and other nodes in the subgraph.
The color of nodes indicate the reciprocal Euclidean distance to the identified node. 

From Fig.~\ref{fig:casestudy}, we observe the following fraud patterns that can be observed on many other nodes in FinGraph:
\begin{enumerate}
	\item Other fraudsters can be found within the 2-hop subgraph of a fraudster, while the 2-hop subgraph of a normal node generally contains other normal users.

	\item Nodes with high reciprocal Euclidean distance (in dark blue) to the identified fraud user at the center of the subgraph may not be fraud users. 
	
	\item Edges connected to fraudsters have TS values spanning a short TS period, while edges connected to normal users do not, showing that fraudsters perform operations for perpetrating frauds in a relatively short time period.
	
	\item Type numbers of edges connected to fraudsters increase as TS values increase, showing that the behaviors of fraud users in FinGraph have certain patterns.
	
\end{enumerate}

\section{Conclusion}

In this paper, we present \model for detecting crowdsourcing fraud. 
We adopt the idea of data augmentation, multi-view learning and contrastive learning when designing \model.
\model can capture the heterogeneity and the dynamics of MMMA data from different views and alleviate label reliance.
Experiments demonstrates the effectiveness of \model.
In the future, we plan to further improve the design of \model by leveraging the fraud patterns that \model helps to discover.

\section*{Acknowledgments}

Hui Li is supported by National Science and Technology Major Project (No. 2022ZD0118201), National Natural Science Foundation of China (No. 62002303, 42171456) and 2021 Tencent WeChat Rhino-Bird Focused Research Program. Jieming Shi is supported by Hong Kong RGC ECS (No. 25201221) and Natural Science Foundation of China (No. 62202404).

\bibliographystyle{ACM-Reference-Format}
\bibliography{main}


\begin{thebibliography}{26}


\ifx \showCODEN    \undefined \def \showCODEN     #1{\unskip}     \fi
\ifx \showDOI      \undefined \def \showDOI       #1{#1}\fi
\ifx \showISBNx    \undefined \def \showISBNx     #1{\unskip}     \fi
\ifx \showISBNxiii \undefined \def \showISBNxiii  #1{\unskip}     \fi
\ifx \showISSN     \undefined \def \showISSN      #1{\unskip}     \fi
\ifx \showLCCN     \undefined \def \showLCCN      #1{\unskip}     \fi
\ifx \shownote     \undefined \def \shownote      #1{#1}          \fi
\ifx \showarticletitle \undefined \def \showarticletitle #1{#1}   \fi
\ifx \showURL      \undefined \def \showURL       {\relax}        \fi
\providecommand\bibfield[2]{#2}
\providecommand\bibinfo[2]{#2}
\providecommand\natexlab[1]{#1}
\providecommand\showeprint[2][]{arXiv:#2}

\bibitem[Ala'raj et~al\mbox{.}(2021)]%
        {AlarajAM21}
\bibfield{author}{\bibinfo{person}{Maher Ala'raj}, \bibinfo{person}{Maysam~F.
  Abbod}, {and} \bibinfo{person}{Munir Majdalawieh}.}
  \bibinfo{year}{2021}\natexlab{}.
\newblock \showarticletitle{Modelling customers credit card behaviour using
  bidirectional {LSTM} neural networks}.
\newblock \bibinfo{journal}{\emph{J. Big Data}} \bibinfo{volume}{8},
  \bibinfo{number}{1} (\bibinfo{year}{2021}), \bibinfo{pages}{69}.
\newblock


\bibitem[Cai et~al\mbox{.}(2021)]%
        {CaiCLGN0C21}
\bibfield{author}{\bibinfo{person}{Lei Cai}, \bibinfo{person}{Zhengzhang Chen},
  \bibinfo{person}{Chen Luo}, \bibinfo{person}{Jiaping Gui},
  \bibinfo{person}{Jingchao Ni}, \bibinfo{person}{Ding Li}, {and}
  \bibinfo{person}{Haifeng Chen}.} \bibinfo{year}{2021}\natexlab{}.
\newblock \showarticletitle{Structural Temporal Graph Neural Networks for
  Anomaly Detection in Dynamic Graphs}. In \bibinfo{booktitle}{\emph{{CIKM}}}.
  \bibinfo{pages}{3747--3756}.
\newblock


\bibitem[Chen and Guestrin(2016)]%
        {ChenG16}
\bibfield{author}{\bibinfo{person}{Tianqi Chen} {and} \bibinfo{person}{Carlos
  Guestrin}.} \bibinfo{year}{2016}\natexlab{}.
\newblock \showarticletitle{XGBoost: {A} Scalable Tree Boosting System}. In
  \bibinfo{booktitle}{\emph{{KDD}}}. \bibinfo{pages}{785--794}.
\newblock


\bibitem[Ding et~al\mbox{.}(2019)]%
        {DingLBL19}
\bibfield{author}{\bibinfo{person}{Kaize Ding}, \bibinfo{person}{Jundong Li},
  \bibinfo{person}{Rohit Bhanushali}, {and} \bibinfo{person}{Huan Liu}.}
  \bibinfo{year}{2019}\natexlab{}.
\newblock \showarticletitle{Deep Anomaly Detection on Attributed Networks}. In
  \bibinfo{booktitle}{\emph{{SDM}}}. \bibinfo{pages}{594--602}.
\newblock


\bibitem[Dou et~al\mbox{.}(2020)]%
        {DouL0DPY20}
\bibfield{author}{\bibinfo{person}{Yingtong Dou}, \bibinfo{person}{Zhiwei Liu},
  \bibinfo{person}{Li Sun}, \bibinfo{person}{Yutong Deng}, \bibinfo{person}{Hao
  Peng}, {and} \bibinfo{person}{Philip~S. Yu}.}
  \bibinfo{year}{2020}\natexlab{}.
\newblock \showarticletitle{Enhancing Graph Neural Network-based Fraud
  Detectors against Camouflaged Fraudsters}. In
  \bibinfo{booktitle}{\emph{{CIKM}}}. \bibinfo{pages}{315--324}.
\newblock


\bibitem[Guardian(2013)]%
        {clickfarm}
\bibfield{author}{\bibinfo{person}{The Guardian}.}
  \bibinfo{year}{2013}\natexlab{}.
\newblock \bibinfo{title}{How low-paid workers at 'click farms' create
  appearance of online popularity}.
\newblock
  \bibinfo{howpublished}{\url{https://www.theguardian.com/technology/2013/aug/02/click-farms-appearance-online-popularity}}.
\newblock
\newblock
\shownote{Accessed Oct 13, 2022}.


\bibitem[Kipf and Welling(2017)]%
        {KipfW17}
\bibfield{author}{\bibinfo{person}{Thomas~N. Kipf} {and} \bibinfo{person}{Max
  Welling}.} \bibinfo{year}{2017}\natexlab{}.
\newblock \showarticletitle{Semi-Supervised Classification with Graph
  Convolutional Networks}. In \bibinfo{booktitle}{\emph{{ICLR}}}.
\newblock
\newblock
\shownote{\url{https://openreview.net/forum?id=SJU4ayYgl}}.


\bibitem[Li et~al\mbox{.}(2014)]%
        {LiSCGY14}
\bibfield{author}{\bibinfo{person}{Nan Li}, \bibinfo{person}{Huan Sun},
  \bibinfo{person}{Kyle~C. Chipman}, \bibinfo{person}{Jemin George}, {and}
  \bibinfo{person}{Xifeng Yan}.} \bibinfo{year}{2014}\natexlab{}.
\newblock \showarticletitle{A Probabilistic Approach to Uncovering Attributed
  Graph Anomalies}. In \bibinfo{booktitle}{\emph{{SDM}}}.
  \bibinfo{pages}{82--90}.
\newblock


\bibitem[Li and Chen(2020)]%
        {li2020comparative}
\bibfield{author}{\bibinfo{person}{Yiheng Li} {and} \bibinfo{person}{Weidong
  Chen}.} \bibinfo{year}{2020}\natexlab{}.
\newblock \showarticletitle{A comparative performance assessment of ensemble
  learning for credit scoring}.
\newblock \bibinfo{journal}{\emph{Mathematics}} \bibinfo{volume}{8},
  \bibinfo{number}{10} (\bibinfo{year}{2020}), \bibinfo{pages}{1756}.
\newblock


\bibitem[Liu et~al\mbox{.}(2021)]%
        {LiuAQCFYH21}
\bibfield{author}{\bibinfo{person}{Yang Liu}, \bibinfo{person}{Xiang Ao},
  \bibinfo{person}{Zidi Qin}, \bibinfo{person}{Jianfeng Chi},
  \bibinfo{person}{Jinghua Feng}, \bibinfo{person}{Hao Yang}, {and}
  \bibinfo{person}{Qing He}.} \bibinfo{year}{2021}\natexlab{}.
\newblock \showarticletitle{Pick and Choose: {A} GNN-based Imbalanced Learning
  Approach for Fraud Detection}. In \bibinfo{booktitle}{\emph{{WWW}}}.
  \bibinfo{pages}{3168--3177}.
\newblock


\bibitem[Liu et~al\mbox{.}(2019)]%
        {LiuCLZLSQ19}
\bibfield{author}{\bibinfo{person}{Ziqi Liu}, \bibinfo{person}{Chaochao Chen},
  \bibinfo{person}{Longfei Li}, \bibinfo{person}{Jun Zhou},
  \bibinfo{person}{Xiaolong Li}, \bibinfo{person}{Le Song}, {and}
  \bibinfo{person}{Yuan Qi}.} \bibinfo{year}{2019}\natexlab{}.
\newblock \showarticletitle{GeniePath: Graph Neural Networks with Adaptive
  Receptive Paths}. In \bibinfo{booktitle}{\emph{{AAAI}}}.
  \bibinfo{pages}{4424--4431}.
\newblock


\bibitem[Lv et~al\mbox{.}(2021)]%
        {LvDLCFHZJDT21}
\bibfield{author}{\bibinfo{person}{Qingsong Lv}, \bibinfo{person}{Ming Ding},
  \bibinfo{person}{Qiang Liu}, \bibinfo{person}{Yuxiang Chen},
  \bibinfo{person}{Wenzheng Feng}, \bibinfo{person}{Siming He},
  \bibinfo{person}{Chang Zhou}, \bibinfo{person}{Jianguo Jiang},
  \bibinfo{person}{Yuxiao Dong}, {and} \bibinfo{person}{Jie Tang}.}
  \bibinfo{year}{2021}\natexlab{}.
\newblock \showarticletitle{Are we really making much progress?: Revisiting,
  benchmarking and refining heterogeneous graph neural networks}. In
  \bibinfo{booktitle}{\emph{{KDD}}}. \bibinfo{pages}{1150--1160}.
\newblock


\bibitem[Ma et~al\mbox{.}(2023)]%
        {9565320}
\bibfield{author}{\bibinfo{person}{Xiaoxiao Ma}, \bibinfo{person}{Jia Wu},
  \bibinfo{person}{Shan Xue}, \bibinfo{person}{Jian Yang},
  \bibinfo{person}{Chuan Zhou}, \bibinfo{person}{Quan~Z. Sheng},
  \bibinfo{person}{Hui Xiong}, {and} \bibinfo{person}{Leman Akoglu}.}
  \bibinfo{year}{2023}\natexlab{}.
\newblock \showarticletitle{A Comprehensive Survey on Graph Anomaly Detection
  With Deep Learning}.
\newblock \bibinfo{journal}{\emph{{IEEE} Trans. Knowl. Data Eng.}}
  \bibinfo{volume}{35}, \bibinfo{number}{12} (\bibinfo{year}{2023}),
  \bibinfo{pages}{12012--12038}.
\newblock


\bibitem[Manzoor et~al\mbox{.}(2016)]%
        {ManzoorMA16}
\bibfield{author}{\bibinfo{person}{Emaad~A. Manzoor},
  \bibinfo{person}{Sadegh~M. Milajerdi}, {and} \bibinfo{person}{Leman Akoglu}.}
  \bibinfo{year}{2016}\natexlab{}.
\newblock \showarticletitle{Fast Memory-efficient Anomaly Detection in
  Streaming Heterogeneous Graphs}. In \bibinfo{booktitle}{\emph{{KDD}}}.
  \bibinfo{pages}{1035--1044}.
\newblock


\bibitem[Massey(1951)]%
        {2280095}
\bibfield{author}{\bibinfo{person}{Frank~J. Massey}.}
  \bibinfo{year}{1951}\natexlab{}.
\newblock \showarticletitle{The Kolmogorov-Smirnov Test for Goodness of Fit}.
\newblock \bibinfo{journal}{\emph{J. Amer. Statist. Assoc.}}
  \bibinfo{volume}{46}, \bibinfo{number}{253} (\bibinfo{year}{1951}),
  \bibinfo{pages}{68--78}.
\newblock


\bibitem[Peng et~al\mbox{.}(2022)]%
        {9162509}
\bibfield{author}{\bibinfo{person}{Zhen Peng}, \bibinfo{person}{Minnan Luo},
  \bibinfo{person}{Jundong Li}, \bibinfo{person}{Luguo Xue}, {and}
  \bibinfo{person}{Qinghua Zheng}.} \bibinfo{year}{2022}\natexlab{}.
\newblock \showarticletitle{A Deep Multi-View Framework for Anomaly Detection
  on Attributed Networks}.
\newblock \bibinfo{journal}{\emph{{IEEE} Trans. Knowl. Data Eng.}}
  \bibinfo{volume}{34}, \bibinfo{number}{6} (\bibinfo{year}{2022}),
  \bibinfo{pages}{2539--2552}.
\newblock


\bibitem[Schlichtkrull et~al\mbox{.}(2018)]%
        {SchlichtkrullKB18}
\bibfield{author}{\bibinfo{person}{Michael~Sejr Schlichtkrull},
  \bibinfo{person}{Thomas~N. Kipf}, \bibinfo{person}{Peter Bloem},
  \bibinfo{person}{Rianne van~den Berg}, \bibinfo{person}{Ivan Titov}, {and}
  \bibinfo{person}{Max Welling}.} \bibinfo{year}{2018}\natexlab{}.
\newblock \showarticletitle{Modeling Relational Data with Graph Convolutional
  Networks}. In \bibinfo{booktitle}{\emph{{ESWC}}}. \bibinfo{pages}{593--607}.
\newblock


\bibitem[Sricharan and Das(2014)]%
        {SricharanD14}
\bibfield{author}{\bibinfo{person}{Kumar Sricharan} {and}
  \bibinfo{person}{Kamalika Das}.} \bibinfo{year}{2014}\natexlab{}.
\newblock \showarticletitle{Localizing anomalous changes in time-evolving
  graphs}. In \bibinfo{booktitle}{\emph{{SIGMOD}}}.
  \bibinfo{pages}{1347--1358}.
\newblock


\bibitem[Tencent(2022)]%
        {wechatuser}
\bibfield{author}{\bibinfo{person}{Tencent}.} \bibinfo{year}{2022}\natexlab{}.
\newblock \bibinfo{title}{Tencent Announces 2022 Second Quarter Results and
  Interim Results}.
\newblock
  \bibinfo{howpublished}{\url{https://static.www.tencent.com/uploads/2022/08/17/a1a39b69021bb7e4bf7f8dd238070079.pdf}}.
\newblock
\newblock
\shownote{Accessed Oct 13, 2022}.


\bibitem[van~der Maaten and Hinton(2008)]%
        {MaatenH08}
\bibfield{author}{\bibinfo{person}{Laurens van~der Maaten} {and}
  \bibinfo{person}{Geoffrey Hinton}.} \bibinfo{year}{2008}\natexlab{}.
\newblock \showarticletitle{Visualizing Data using t-SNE}.
\newblock \bibinfo{journal}{\emph{J. Mach. Learn. Res.}}  \bibinfo{volume}{9}
  (\bibinfo{year}{2008}), \bibinfo{pages}{2579--2605}.
\newblock


\bibitem[Velickovic et~al\mbox{.}(2018)]%
        {VelickovicCCRLB18}
\bibfield{author}{\bibinfo{person}{Petar Velickovic}, \bibinfo{person}{Guillem
  Cucurull}, \bibinfo{person}{Arantxa Casanova}, \bibinfo{person}{Adriana
  Romero}, \bibinfo{person}{Pietro Li{\`{o}}}, {and} \bibinfo{person}{Yoshua
  Bengio}.} \bibinfo{year}{2018}\natexlab{}.
\newblock \showarticletitle{Graph Attention Networks}. In
  \bibinfo{booktitle}{\emph{{ICLR}}}.
\newblock
\newblock
\shownote{\url{https://openreview.net/forum?id=rJXMpikCZ}}.


\bibitem[Wang et~al\mbox{.}(2021)]%
        {Wang0GYL021}
\bibfield{author}{\bibinfo{person}{Yanling Wang}, \bibinfo{person}{Jing Zhang},
  \bibinfo{person}{Shasha Guo}, \bibinfo{person}{Hongzhi Yin},
  \bibinfo{person}{Cuiping Li}, {and} \bibinfo{person}{Hong Chen}.}
  \bibinfo{year}{2021}\natexlab{}.
\newblock \showarticletitle{Decoupling Representation Learning and
  Classification for GNN-based Anomaly Detection}. In
  \bibinfo{booktitle}{\emph{{SIGIR}}}. \bibinfo{pages}{1239--1248}.
\newblock


\bibitem[Wu et~al\mbox{.}(2023)]%
        {abs-2105-07342}
\bibfield{author}{\bibinfo{person}{Lirong Wu}, \bibinfo{person}{Haitao Lin},
  \bibinfo{person}{Cheng Tan}, \bibinfo{person}{Zhangyang Gao}, {and}
  \bibinfo{person}{Stan~Z. Li}.} \bibinfo{year}{2023}\natexlab{}.
\newblock \showarticletitle{Self-Supervised Learning on Graphs: Contrastive,
  Generative, or Predictive}.
\newblock \bibinfo{journal}{\emph{{IEEE} Trans. Knowl. Data Eng.}}
  \bibinfo{volume}{35}, \bibinfo{number}{4} (\bibinfo{year}{2023}),
  \bibinfo{pages}{4216--4235}.
\newblock


\bibitem[You et~al\mbox{.}(2022)]%
        {YouDL22}
\bibfield{author}{\bibinfo{person}{Jiaxuan You}, \bibinfo{person}{Tianyu Du},
  {and} \bibinfo{person}{Jure Leskovec}.} \bibinfo{year}{2022}\natexlab{}.
\newblock \showarticletitle{{ROLAND:} Graph Learning Framework for Dynamic
  Graphs}. In \bibinfo{booktitle}{\emph{{KDD}}}. \bibinfo{pages}{2358--2366}.
\newblock


\bibitem[Yu et~al\mbox{.}(2018)]%
        {YuCAZCW18}
\bibfield{author}{\bibinfo{person}{Wenchao Yu}, \bibinfo{person}{Wei Cheng},
  \bibinfo{person}{Charu~C. Aggarwal}, \bibinfo{person}{Kai Zhang},
  \bibinfo{person}{Haifeng Chen}, {and} \bibinfo{person}{Wei Wang}.}
  \bibinfo{year}{2018}\natexlab{}.
\newblock \showarticletitle{NetWalk: {A} Flexible Deep Embedding Approach for
  Anomaly Detection in Dynamic Networks}. In \bibinfo{booktitle}{\emph{{KDD}}}.
  \bibinfo{pages}{2672--2681}.
\newblock


\bibitem[Zheng et~al\mbox{.}(2019)]%
        {ZhengLLLG19}
\bibfield{author}{\bibinfo{person}{Li Zheng}, \bibinfo{person}{Zhenpeng Li},
  \bibinfo{person}{Jian Li}, \bibinfo{person}{Zhao Li}, {and}
  \bibinfo{person}{Jun Gao}.} \bibinfo{year}{2019}\natexlab{}.
\newblock \showarticletitle{AddGraph: Anomaly Detection in Dynamic Graph Using
  Attention-based Temporal {GCN}}. In \bibinfo{booktitle}{\emph{{IJCAI}}}.
  \bibinfo{pages}{4419--4425}.
\newblock


\end{thebibliography}

\appendix

\section{Model Size}
\label{app:modelsize}

Let $R$ and $f$ be the number of relation types and the dimensionality of representations.
We provide the analysis of the model size of \app in the following:

\subsection{Model size of HG-Encoder}

The model parameters of one GNN layer in HG-Encoder are $\mathbf{W_{\mathbf{m}}}\in\mathbb{R}^{R\times f\times f}$ and $\mathbf{b_\mathbf{g}}\in\mathbb{R}^{R\times f}$ in Eq.~2, $\mathbf{W_{\mathbf{s}}}\in\mathbb{R}^{R\times f\times f}$ and $\mathbf{b_\mathbf{s}}\in\mathbb{R}^{R\times f}$ in Eq.~3, $\mathbf{W_{\mathbf{q}}}\in\mathbb{R}^{R\times f\times 2f}$ and $\mathbf{b_\mathbf{q}}\in\mathbb{R}^{R\times 2f}$ in Eq.~4, and $\mathbf{W_{\mathbf{w}}}\in\mathbb{R}^{R\times f\times f}$,  $\mathbf{b_\mathbf{w}}\in\mathbb{R}^{R\times f}$ and $\mathbf{a}\in\mathbb{R}^{f\times1}$ in Eq.~5 for $R$ relations in HTG. 
Therefore, the size of parameters of one GNN layer in HG-Encoder is $(5Rf+5R+1)f$.

The prediction layer of HG-Encoder consists of a linear mapping layer followed by a sigmoid function, which is parameterized by $\mathbf{W_{\mathbf{p}}}\in\mathbb{R}^{f\times 1}$ and  $\mathbf{b_\mathbf{p}}\in\mathbb{R}^{f\times 1}$.

Denote the number of GNN layers as $l$ (we use $l=2$). 
The model size of HG-Encoder is $\mathcal{S}_{\text{HG}}=(5l\cdot Rf+5l\cdot R+l+2)f$.
Note that $R$ is 7 and 11 for \app and FinGraph datasets, respectively.
Since $l$ and $R$ are much smaller than $f$ (we set $f=64$), $\mathcal{S}_{\text{HG}}$ is in the magnitude of $f^2$.

\subsection{Model size of CS-Encoder}

The CS-Encoder mainly consists of a position encoding matrix $\mathbf{P}\in\mathbb{R}^{T\times f}$ and the multi-head self-attention module parameterized by matrix $\mathbf{W_{\mathbf{Q}}}\in\mathbb{R}^{h\times f\times f'}$,  $\mathbf{W_{\mathbf{K}}}\in\mathbb{R}^{h\times f\times f'}$ and $\mathbf{W_{\mathbf{V}}}\in\mathbb{R}^{h\times f\times f'}$, where $h$ is the number of heads and $f'$ is the hidden state size of the encoding block. Assume we stack $l_{\text{CS}}$ encoding blocks to construct the CS-Encoder, Then the model size of CS-Encoder can be estimated as $\mathcal{S}_{\text{CS}}=l_{\text{CS}}\cdot(T+3h\cdot f')f$.
In our experiments, we use $l_{\text{CS}}=2$, $T=14$, $h=8$, and $f'$ is set as the same as $f$.
Hence, $\mathcal{S}_{\text{CS}}$ is in the magnitude of $f^2$.

\subsection{Model size of HG-Encoder$_{\text{detect}}$}

HG-Encoder$_{\text{detect}}$ uses the identical design as HG-Encoder. Hence, the model size of HG-Encoder$_{\text{detect}}$ is $\mathcal{S}_{\text{HG}}$.

\subsection{Model size of the complete CMT}

Sum up model size of HG-Encoder, TSS-Encoder (it contains HG-Encoder$_\text{TSS}$ and a CS-Encoder), URS-Encoder (most of it is a CS-Encoder), and HG-Encoder$_{\text{detect}}$, the model size of CMT is roughly $\mathcal{S}_{\text{CMT}} = 3\mathcal{S}_{\text{HG}} + 2\mathcal{S}_{\text{CS}}$.
We can see that the model size of CMT is in the magnitude of $f^2$.

\section{Additional Experiments}
\label{app:exp}

The representations generated by the pretraining phase of \model can be utilized as additional input node features to other detection methods and improve the detection. 
Tab.~\ref{tab:xgb} provides the result of enhancing XGBoost.

\begin{table}[t]
\centering
\caption{Improvements when using pretrained representations to improve XGBoost.}
\label{tab:xgb}
\scalebox{0.9}{
\begin{tabular}{|c|cc|cc|}
\hline
                                                                              & \multicolumn{2}{c|}{WeChat}            & \multicolumn{2}{c|}{FinGraph}          \\ \hline
XGBoost                                                                       & \multicolumn{1}{c|}{AUC}     & KS      & \multicolumn{1}{c|}{AUC}     & KS      \\ \hline
Original features                                                                      & \multicolumn{1}{c|}{0.7189}  & 0.3281  & \multicolumn{1}{c|}{0.7388}  & 0.3911  \\ \hline
\begin{tabular}[c]{@{}c@{}}Combined additional features\\and original features\end{tabular} & \multicolumn{1}{c|}{0.8622}  & 0.5751  & \multicolumn{1}{c|}{0.8245}  & 0.5364  \\ \hline
Improve.\%                                                                    & \multicolumn{1}{c|}{+19.9\%} & +75.3\% & \multicolumn{1}{c|}{+11.6\%} & +37.2\% \\ \hline
\end{tabular}
}
\end{table}

\end{document}